\documentclass{emulateapj}
\usepackage{natbib}
\usepackage{lscape}

\def\gsim{\;\rlap{\lower 2.5pt
 \hbox{$\sim$}}\raise 1.5pt\hbox{$>$}\;}
\def\lsim{\;\rlap{\lower 2.5pt
   \hbox{$\sim$}}\raise 1.5pt\hbox{$<$}\;}

\def\msolarunits{$h^{-1}$ $M_{\odot}$ }
\def\lengthunits{$h^{-1}$ kpc }

\begin{document}

\title{The Self--Regulated Growth of Supermassive Black Holes}
\author{Joshua D. Younger\altaffilmark{1,2}, Philip F. Hopkins\altaffilmark{1}, T. J. Cox\altaffilmark{1,3}, \& Lars Hernquist\altaffilmark{1}}
\altaffiltext{1}{Harvard--Smithsonian Center for Astrophysics, 60 Garden Street, Cambridge, MA 02138}
\altaffiltext{2}{jyounger@cfa.harvard.edu}
\altaffiltext{3}{Keck Foundation Fellow}

\begin{abstract}

We present a series of simulations of the self--regulated growth of
supermassive black holes (SMBHs) in galaxies via three different
fueling mechanisms: major mergers, minor mergers, and disk
instabilities.  The SMBHs in all three scenarios follow the same black
hole fundamental plane (BHFP) and correlation with bulge binding
energy seen in simulations of major mergers, and observed locally.
Furthermore, provided that the total gas supply is significantly
larger than the mass of the SMBH, its limiting mass is not influenced
by the amount of gas available or the efficiency of black hole growth.
This supports the assertion that SMBHs accrete until they reach a
critical mass at which feedback is sufficient to unbind the gas
locally, terminating the inflow and stalling further growth.  At the
same time, while minor and major mergers follow the same projected
correlations (e.g., the $M_{BH}-\sigma$ and Magorrian relations),
SMBHs grown via disk instabilities do not, owing to structural
differences between the host bulges.  This finding is supported by
recent observations of SMBHs in pseudobulges and bulges in barred
systems, as compared to those hosted by classical bulges.  Taken
together, this provides support for the BHFP and binding energy
correlations as being more ``fundamental" than other proposed
correlations in that they reflect the physical mechanism driving the
co-evolution of SMBHs and spheroids.

\end{abstract}


\section{Introduction}
\label{sec:intro}

Over the past two decades, it has been established that supermassive
black holes (SMBHs) in the nuclei of galaxies are common, and that
their masses are connected to the properties of their hosts.  As the
sample of robust SMBH measurements has increased
\citep{kormendy1995,kormendy1994} several black hole - host galaxy
correlations \citep[see][for a review]{novak2006} have
been proposed, including central velocity dispersion
\citep{ferrarese2000,gebhardt2000,tremaine2002}, bulge mass/luminosity
\citep{kormendy1995,magorrian1998,mclure2002,marconi2003,haring2004},
light concentration \citep{graham2001,graham2007}, and binding energy
\citep{aller2007}.  Recently, it has been suggested that these
observed correlations may be projections of a ``Black--Hole
Fundamental Plane"
\citep[BHFP:][]{hopkins2007theory,hopkins2007obs,aller2007,barway2007},
analogous to that describing the structural properties of elliptical
galaxies \citep{dressler1987,djorgovski1987}, that is manifest as a
tilted bulge binding energy correlation.  These relationships are
indicative of an intimate connection between SMBH growth and
galaxy formation/evolution.

\begin{figure*}
\plottwo{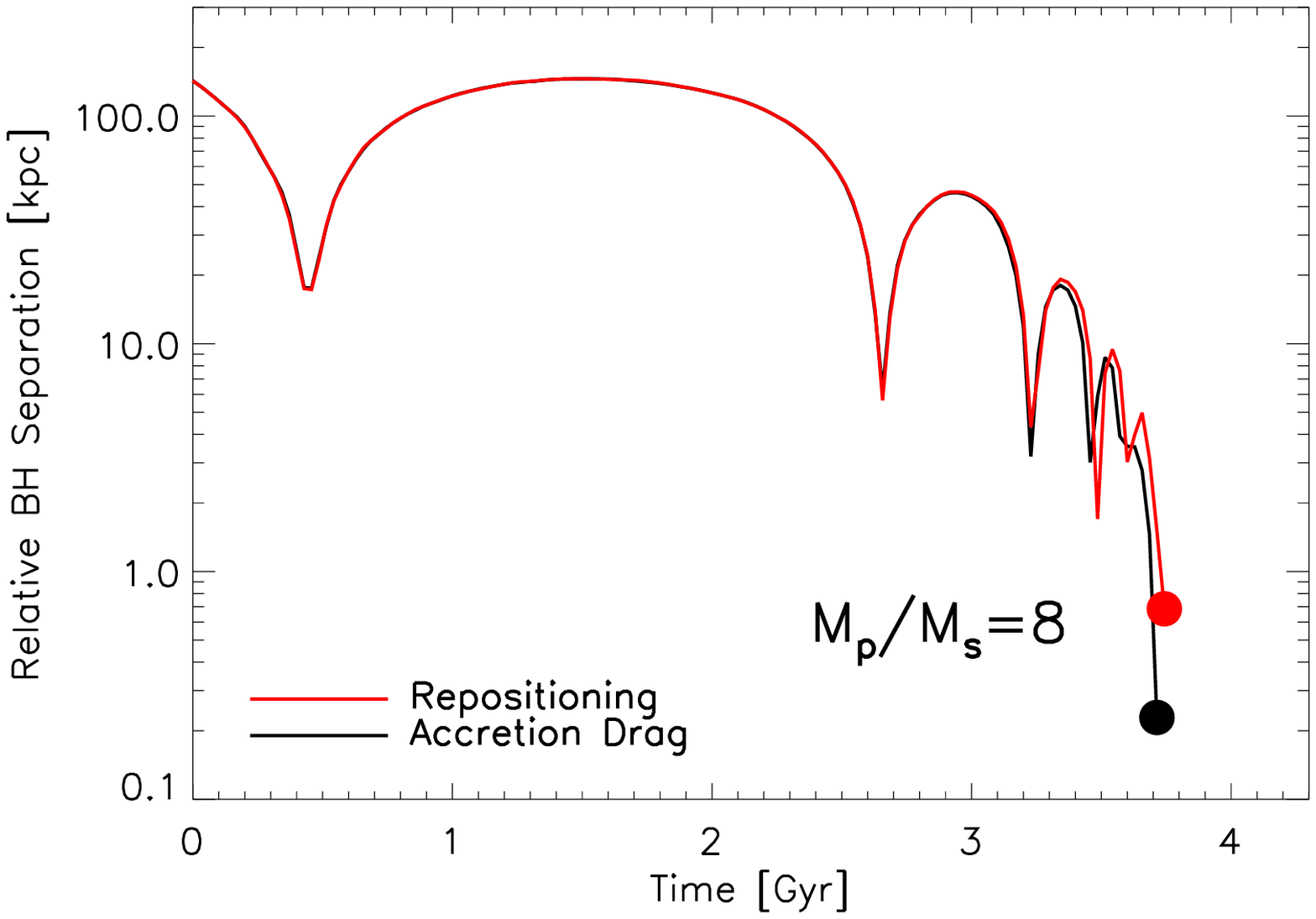}{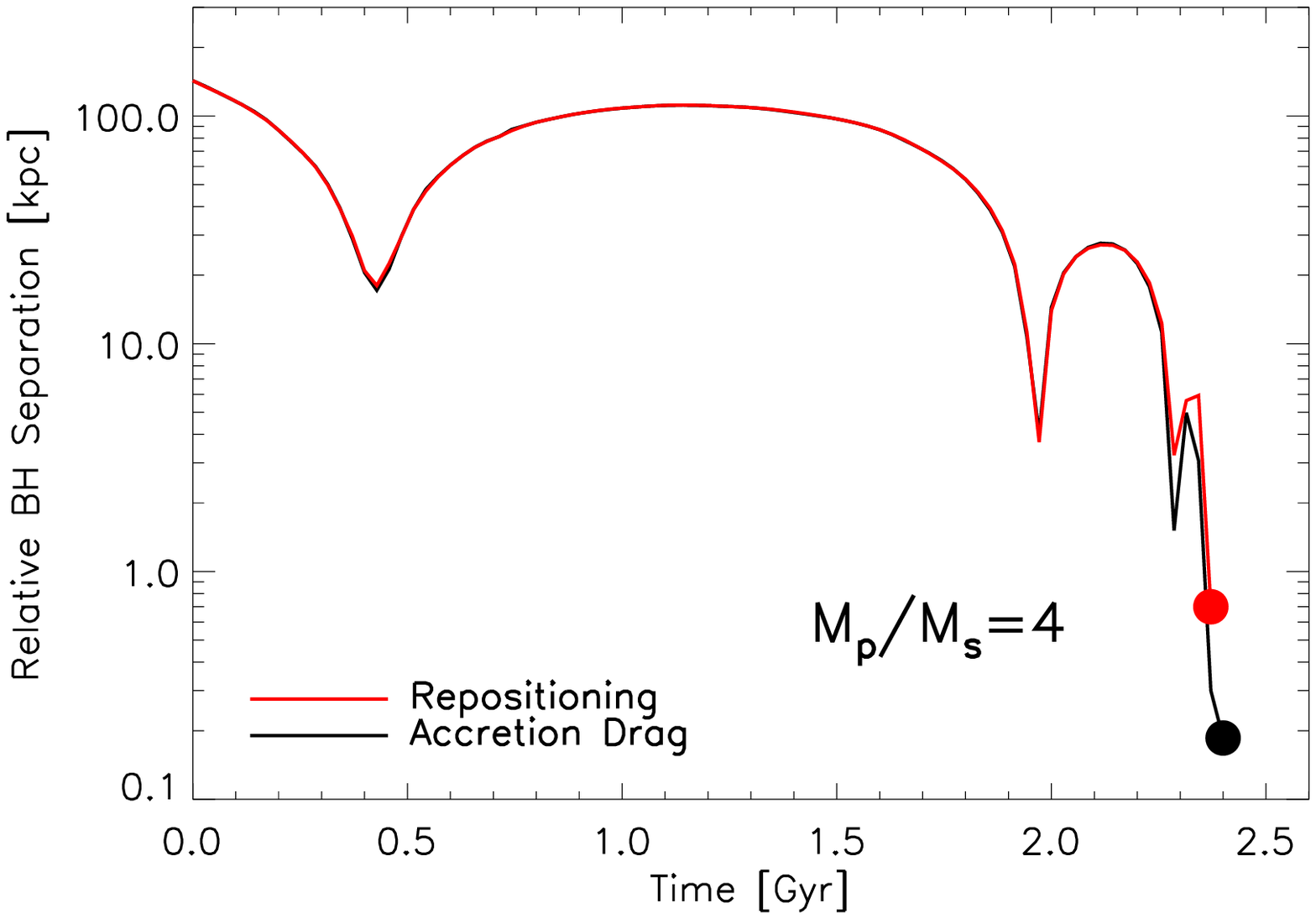}
\plottwo{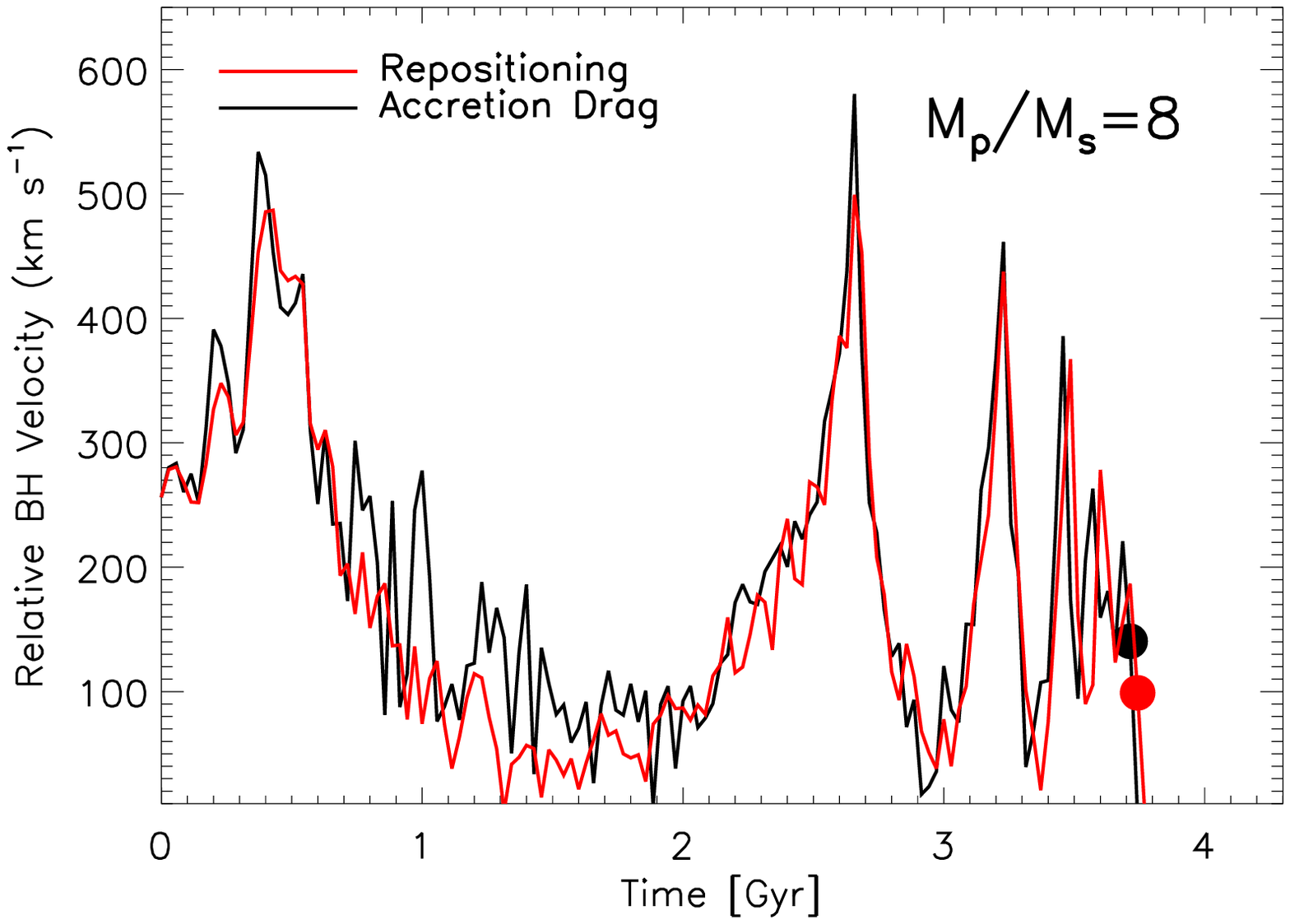}{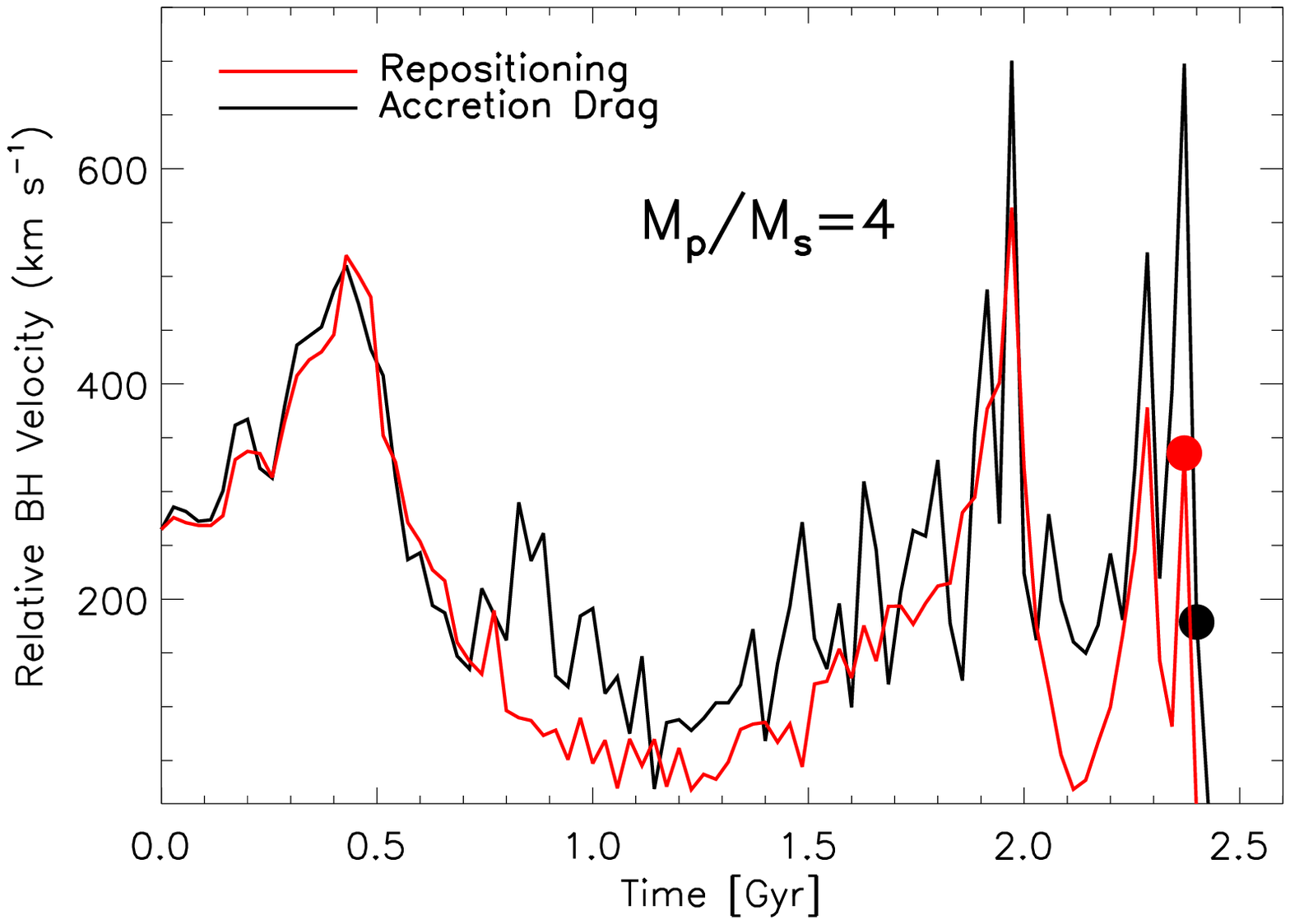}
\caption{{\sc Top:}  Relative separation of the two SMBHs over the course of an $M_p/M_s = 8$ (left) and 4 (right) merger simulation.  Included are results using both the accretion drag (black line; see \S~\ref{sec:bh_mergers} for details) and repositioning \citep[red line; see][]{johansson2008} methods. The circles indicate the time of the merger.  {\sc Bottom:} Same labeling for the relative velocities of the two SMBHs over the course of the simulation.}
\label{fig:bhmerge_relpos}
\end{figure*}

There is also increasing evidence, both direct and indirect, that the
growth of SMBHs is self--regulated, with strong feedback from the
active galactic nucleus (AGN) playing an important role.  Numerical
simulations of major mergers of gas--rich spirals that include this
process, along with gas dissipation, naturally account for
the observed SMBH correlations
\citep{dimatteo2005,robertson2006a,hopkins2007theory,johansson2008},
the fundamental plane of elliptical galaxies \citep{robertson2006b,
hopkins2008e},
the structure \citep{hopkins2008a,hopkins2008b,hopkins2008c} and stellar
kinematics \citep{cox2006} of massive ellipticals, and the redshift
evolution of the correlations \citep{hopkins2007theory,hopkins2008d}.
Moreover, a
strong feedback mode has proven essential in semi--analytic models 
and numerical simulations of
hierarchical galaxy formation \citep[e.g.,][Sommerville et al. 2008,
in prep.]{croton2006,bower2006,sijacki2007,dimatteo2008} 
and the merger--driven evolution of
quasars and red galaxies
\citep{hopkins2005c,hopkins2005b,hopkins2005a,hopkins2005d,hopkins2006a,
hopkins2006b,hopkins2008f,hopkins2008g}.
The prevalence of this process is further supported by direct
observational evidence for outflows in both Seyfert galaxies
\citep[e.g.,][]{crenshaw1999,kriss2000,kaastra2000,kaastra2002,kaspi2000,kaspi2002}
and more luminous quasars
\citep[e.g.,][]{weymann1991,korista1993,arav2001} driven by strong
thermal and radiative feedback originating from the AGN \citep[see
review by][]{crenshaw2003}.

Despite these observational and theoretical advances, the fundamental
character of these SMBH correlations remains poorly understood.  It
appears clear the SMBHs are closely linked with the structural
properties of their host galaxy's bulge \citep[see
e.g.,][]{novak2006}.  However, an important question has not been
adequately addressed: to the extent that the growth of SMBHs is
self--regulated, what property -- or properties -- of this bulge does
the SMBH ``see?"  The more ``fundamental" the scaling relation, the
more it reflects the physical mechanism driving the co-evolution of
SMBHs and bulges.  \citet{hopkins2007theory} suggest that the final
mass of the SMBH is set by the depth of the local potential; the SMBH
grows until feedback unbinds the local gas supply, abruptly
terminating its growth.  However, this hypothesis, while promising,
has not yet been systematically tested.

\begin{figure*}
\plottwo{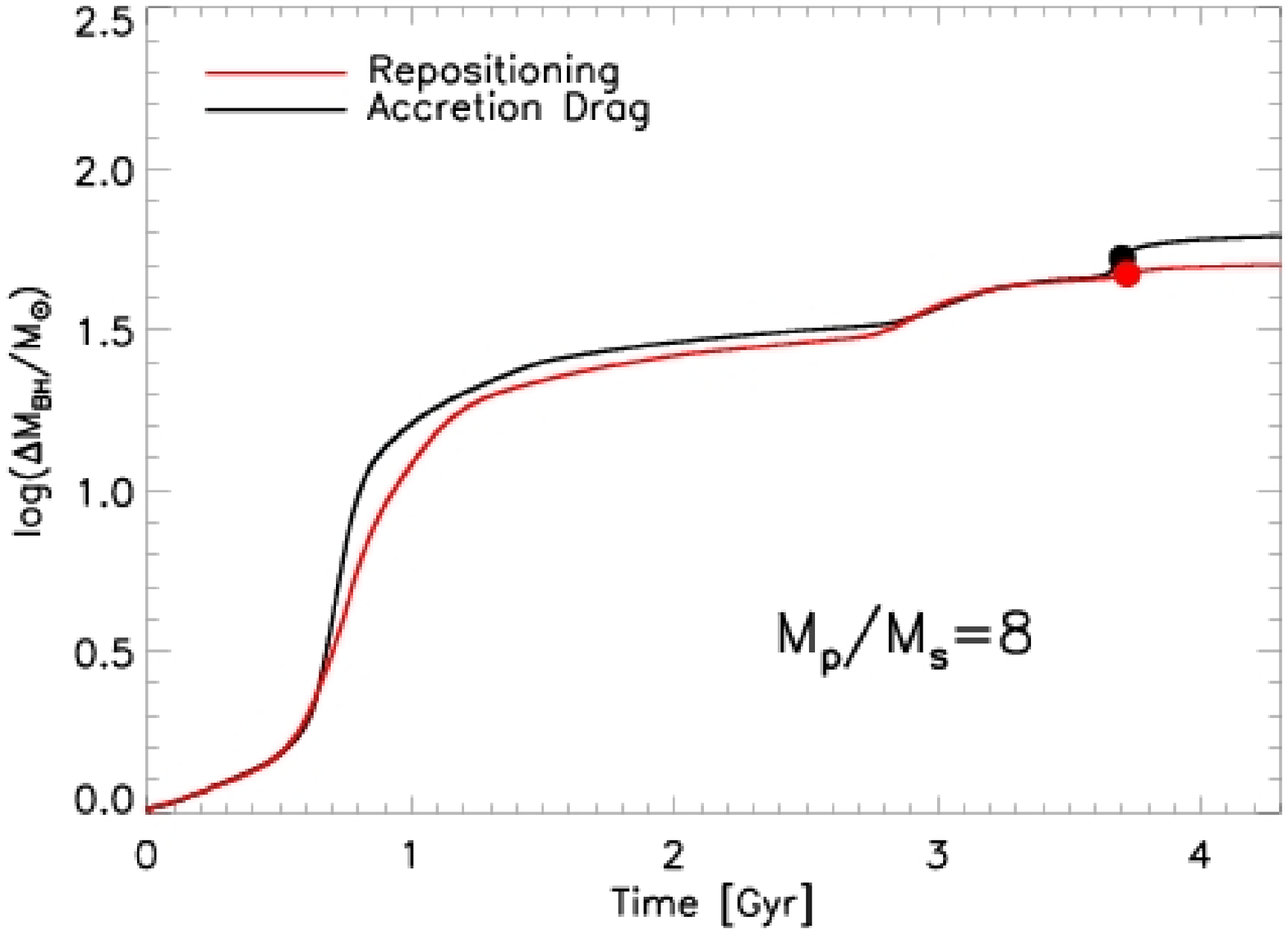}{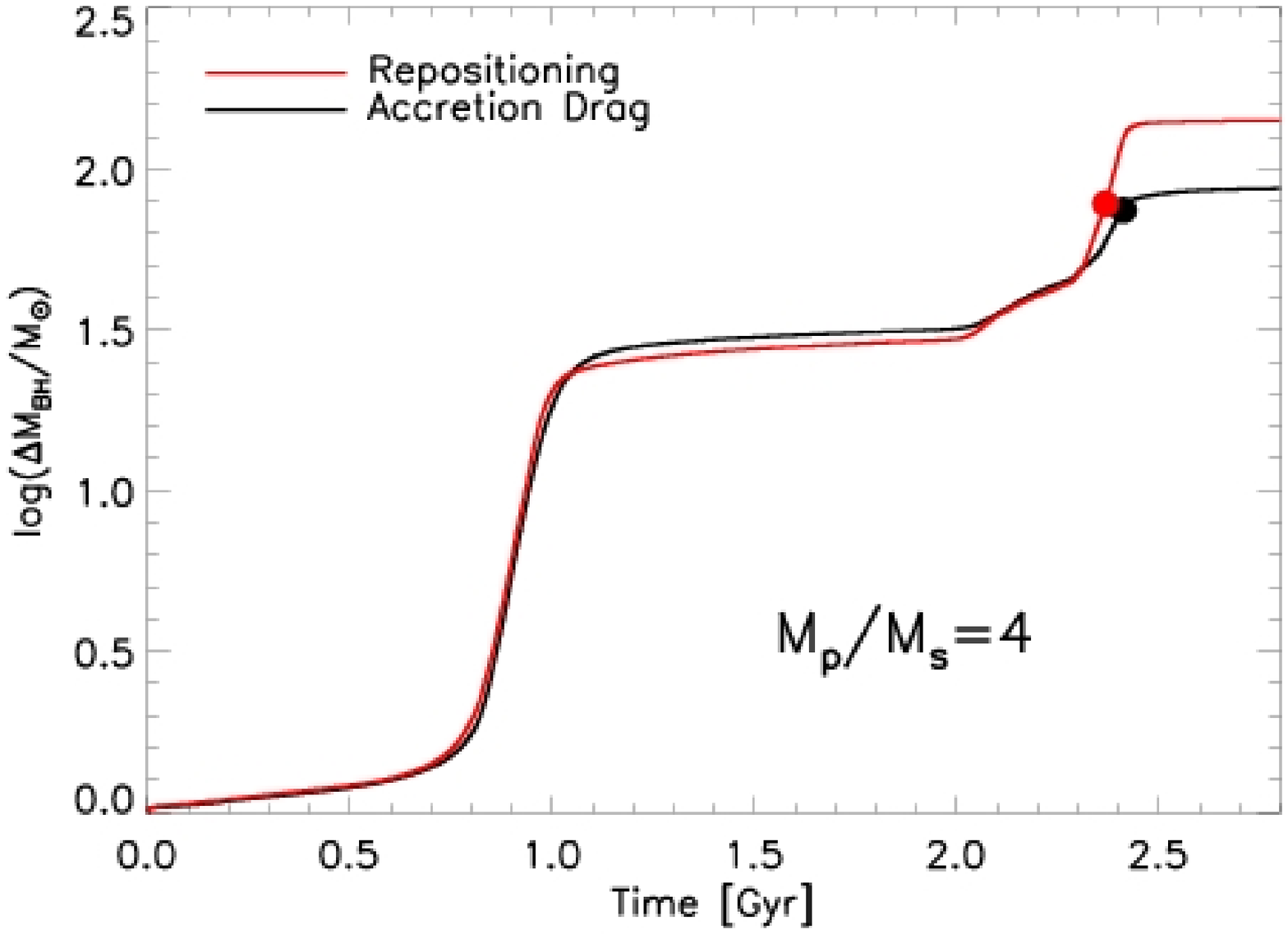}
\caption{Growth of the SMBH over the course of an $M_p/M_s = 8$ (left) and 4 (right) merger simulation.  Included are results using both the accretion drag (black line; see \S~\ref{sec:bh_mergers} for details) and repositioning \citep[red line; see][]{johansson2008} methods. The circles indicate the time of the merger.  There is no systematic trend with the choice of either the accretion drag or repositioning methods.}
\label{fig:bhmerge_grow}
\end{figure*}

In this paper, we use hydrodynamical simulations to examine the
self--regulated growth of SMBHs in three different fueling modes:
major mergers, minor mergers, and disk instabilities.  Our aim is to
investigate the physical mechanism that determines the final mass of
the SMBH.  It is organized as follows: in \S~\ref{sec:methods} we
review our methodology, in \S~\ref{sec:sims} we describe the
simulations, in \S~\ref{sec:results} we present our results, and in
\S~\ref{sec:discussion} we discuss implications.

\section{Methods}
\label{sec:methods}

\subsection{Hydrodynamics}
\label{sec:methdos_hydro}

In what follows, we describe a series of simulations run using {\sc
PGadget2} \citep{springel2005}, a massively parallel smoothed particle
hydrodynamics code \citep[SPH: see also][]{lucy1977,gingold1977} using
the entropy conserving formulation of \citet{springelhernquist2002},
coupled with a Tree algorithm \citep{barnes1986} to compute the
gravitational forces between particles \citep[see
also][]{hernquist1989}.  We also include the effects of radiative
cooling \citep[see][]{katz1996}, and employ a sub--resolution model
for the interstellar medium (ISM) to describe star formation and
supernova feedback, tuned to reproduce the local Schmidt Law
\citep{schmidt1959,kennicutt1998}, that pressurizes the star-forming
gas via an effective equation of state $q_{EOS}$ that interpolates
between an isothermal ($q_{EOS}=0$) and full multiphase ISM
\citep[$q_{EOS}=1$;][]{springelhernquist2003}.  This sub--resolution
model is essential for simulations of gas rich systems, because it
allows us to evolve disks of arbitrary gas content without fragmenting
owing to \citet{toomre1964} instability
\citep[see][]{springelhernquist2003,springelhernquist2005,
robertson2004,robertson2006c}.

\subsection{SMBH Growth and Feedback}
\label{sec:methods_feedback}

SMBHs in our simulations are represented by ``sink" particles, which
increase their mass at the rate $\dot{M}_{BH}$ via Eddington--limited
Bondi--Hoyle--Lyttleton accretion
\citep{hoyle1941,bondi1944,bondi1952} of the surrounding gas.  
As the numerical prefactor for the accretion rate -- Bondi accretion is a dimensionless scaling --
we adopt an efficiency of $\alpha = 100$ for all the simulations used in this work 
\citep[see also][]{SDH2005,johansson2008}.  However, this choice matters little as the 
SMBH's growth is Eddington limited during all the times of interest.  The
bolometric luminosity of the SMBH is assumed to be $L_{bol} = \epsilon
\dot{M} c^2$, where $\epsilon = 0.1$ is the canonical radiative
efficiency for thin--disk accretion.  Feedback is implemented by
coupling a fixed fraction of this luminosity (feedback efficiency
$\eta$; typically $\eta \approx 5\%$) to its environment by injecting
thermal energy into the surrounding gas, weighted by the SPH smoothing
kernel.  The net ratio between the dimensional accretion rate (Bondi or Eddington) and 
the thermal feedback input into the gas is in some sense 
the only free parameter -- $\alpha$, $\epsilon$, and $\eta$ 
are degenerate -- in our feedback
model, and that ratio is chosen to reproduce the normalization of the local
$M_{BH}-\sigma$ relation \citep{dimatteo2005,robertson2006a}; the
slopes and relative normalizations of the resulting correlations of
SMBH mass with galaxy properties are generic, and are generally not
tunable.

It may be argued that this specific method of implementing feedback --
in this case an isotropic thermal coupling -- is significant for our
results, especially in the absence of constraints from a
well--understood, detailed model for AGN feedback.  It is worth noting
that alternate feedback prescriptions have led to similar results as
those presented here \citep[e.g.,][]{murray2005}.  However, more
generally our isotropic thermal coupling to the surrounding gas is a
good approximation to a variety of feedback mechanisms, as it leads
to a shock front which isotropizes and becomes well--mixed over
physical scales smaller than those relevant to our simulations, and
in timescales smaller than the dynamical time of the galaxy
\citep{hopkinshernquist2006,hopkins2006c}.\footnote{This approximation is not valid 
in the case of ``radio--mode" feedback \citep{croton2006}, in which MHD effects can 
maintain a collimated jet on scale comparable to the galaxy.  However, observations suggest 
that black holes gain most of their mass in bright quasar modes \citep{soltan1982}, and $\lsim 10\%$ of 
these objects are radio--loud.  Therefore, we do not believe that this will be the relevant feedback
mechanism for the growth mechanisms presented here.}
The spatial (i.e., SPH smoothing length) resolution for the
simulations analyzed here is $30-50$ pc, and the dynamical time over
those scales is $t_{dyn}({\rm 30-50 pc})\sim 10^7$ years.  Numerical
experiments have shown that shock fronts, even if initially highly
beamed, will become spherical on an isotropization timescale of
$t_{\textrm{\sc iso}} \approx 6\times10^3 (E_{51}/\rho_{24})^{1/3}$ years --
where $E_{51}$ is the input energy in units of $10^{51}$ ergs and
$\rho_{24}$ is the density of the ambient ISM in units of $10^{-24}$ g cm$^{-3}$ -- and size scales of
$\lsim 10$ pc \citep{ayal2001}, both of which are below scales probed
by our simulations.  Therefore, for the simulations presented here, we
believe that an isotropic thermal coupling prescription for feedback
offers a good approximation \citep[for further discussion, see][]{hopkins2006a}; 
an assertion that we will test in detail in due course.

\subsection{Black Hole Mergers}
\label{sec:bh_mergers}

In simulations of interacting systems the implementation of SMBH mergers can 
potentially introduce a systematic bias in the final SMBH mass.   Absent resolution 
constraints, we would expect the BH binary to eventually harden via a combination of 
dynamical \citep{makino2004,berczik2006} and hydrodynamic \citep{escala2004} processes, 
at which point gravitational wave emission dominates the energy loss and the two SMBHs 
in-spiral and coalesce \citep[e.g.,][]{milosavljevic2001}.  However, because computational 
limitations prevent following this evolution in a galaxy--scale simulation -- and 
gravitational radiation is not modeled at all -- the standard prescription adopted by \citet{SDH2005} 
merges the two SMBHs if they are within a smoothing length with a relative velocity less than
or equal to the local sound speed.   

Like \citet{johansson2008}, we found that this simple criterion was not in of itself adequate to 
ensure rapid merging of the SMBHs at the end of the merger -- particularly for the 
unequal mass interactions.  However, rather than employ
their repositioning scheme -- wherein the SMBH is repositioned at every timestep at the minimum of 
the local potential -- we introduced a term accounting for Bondi accretion drag \citep{edgar2004}:
\begin{equation}
F_{drag} = \dot{M}_{BH} v_\infty
\end{equation}
where $v_\infty$ is the initial infall speed of the gas in the Bondi approximation.  This is a very approximate 
treatment \citep[for a more detailed analysis, see][]{ruderman1971}, but captures the essential physics at a level that
is adequate for our purposes.

\begin{table*}
\begin{center}
\caption{Progenitor Disk Structural Parameters }
\begin{tabular}{cccccccc}
\hline
\hline
& $V_{200}$ & $M_{200}$ & $c$ & $h_{D}$ & $N_{halo}$ & $N_{baryons}$ \\
& [km s$^{-1}$] & [$h^{-1} 10^{10} M_\odot$] & & [$h^{-1}$ kpc] & & \\
\hline
{\sc Sb} & 160 & 95 & 9 & 4.1 & $3.3\times 10^5$ & $2.0\times 10^5$ \\
{\sc Sc} & 130 & 51 & 10 & 3.2 & $1.8\times 10^5$ & $1.1\times 10^5$  \\
{\sc Sd} & 100 & 23 & 12 & 2.2 & $8.1\times 10^4$ & $4.9\times 10^4$ \\
{\sc Im} & 80 & 12  & 12 & 1.6 & $4.2\times 10^4$ & $1.5\times 10^4$ \\
\hline
\hline
\end{tabular}
\label{tab:models}
\end{center}
\end{table*}

To investigate the sensitivity of our results to this choice of numerical implementation, we ran two 
representative simulations -- an $M_p/M_s = 8$ and 4 interaction 
with initial gas fractions of $f_g = 0.8$\footnote{This high gas fraction was chosen to ensure that the primary had 
a significant gas fraction -- in this case $f_g\approx 0.3$ -- at the time of the SMBH merger.}
 and 0.4 respectively and identical orbital parameters (see \S~\ref{sec:mergers} for details) -- 
with both accretion drag and repositioning.   In Figure~\ref{fig:bhmerge_relpos} we show the relative separations
and velocities of the two SMBHs, and in Figure~\ref{fig:bhmerge_grow} we show the SMBH growth history.  
There is no apparent systematic trend based on the particularly choice of implementation.

Based on this simple experiment, we do not see a systematic trend in the use of accretion drag over repositioning.  
Rather, the choice of one or the other method introduces a scatter of $\sim 0.1-0.2$ dex, which is noticeably smaller than both the simulation--to--simulation 
scatter, and even the uncertainty associated with the stochastic nature of SMBH accretion.  Therefore, while we note that
a more systematic study of SMBH merger prescriptions for hydrodynamical simulations is warranted, our choice of the 
accretion drag method does not appear to bias our results. 
However, we do prefer it over repositioning for its numerical stability; in some -- admittedly rare -- cases, 
and particularly when the mass ratio of the interaction is high and one galaxy dominates the local potential, 
the central SMBH in the secondary can decouple from its host galaxy and follows the gravitational field lines of 
the primary, resulting in a spurious SMBH merger.  However, in general, seeing as the choice of method does not qualitatively change 
our results, and there is no clear physical argument to favor either, we consider it one of the systematic uncertainties in our modeling.

\subsection{Progenitor Disks}
\label{sec:methods_prog}

The progenitor disk models are constructed following \citet{SDH2005}.
Exponential disks of gas and stars -- and optionally a compact bulge
-- are embedded in a dark matter halo with a \citet{hernquist1990}
density profile, as motivated by cosmological N--body simulations
\citep[e.g.,][]{nfw1996,bullock2001,busha2005}.  Their total mass is
$V_{200}^3/(10 G H_0)$, where $V_{200}$ is the maximum circular
velocity at an overdensity of $200\rho_c$, with a disk mass fraction
of $m_d$, disk gas fraction of $f_g = m_g/(m_d +m_g)$, and a bulge
mass fraction of $m_b$.  The disk scale--length is computed assuming a
spin parameter of $\lambda = 0.033-0.05$, again motivated by
cosmological N--body simulations
\citep[e.g.,][]{cole1996,vitvitska2002,maccio2007,bett2007}.

For this work, we use three sets of simulations: (1) major mergers,
(2) a mass ratio series, and (3) isolated, bar--unstable disks (hereafter
referred to as simply ``unstable disks").  The structural parameters
of the progenitor disks used in the major merger series are discussed
in detail by \citet{robertson2006a,robertson2006b}, and we refer the reader to those
references for the specifics.  The initial disks used in both the
mass ratio and unstable disk series are described in
Table~\ref{tab:models}, and are similar to those used in
\citet{younger2007}.  

\section{Simulations}

\label{sec:sims}

\subsection{Mergers}
\label{sec:mergers}

\begin{table*}
\begin{center}
\caption{Orbital and Initial Disk Parameters for the Mass Ratio Series}
\begin{tabular}{ccccc}
\hline
\hline
& Name &  Mass Ratio & Inclination & Gas Fraction\\
& & $M_p/M_s$ & $i$ & $f_{g,p},f_{g,s}$ \\
\hline
Standard  & {\sc Sb}fg4{\sc Im}fg4 & 8 & $0^\circ,30^\circ,90^\circ,150^\circ,180^\circ$ &  0.4, 0.4 \\
 & {\sc Sb}fg4{\sc Sd}fg4 & 4 & $0^\circ,30^\circ,90^\circ,150^\circ,180^\circ$ &  0.4, 0.4 \\
 & {\sc Sb}fg4{\sc Sc}fg4 & 2 & $0^\circ,30^\circ,90^\circ,150^\circ,180^\circ$ &  0.4, 0.4 \\
 & {\sc Sb}fg4{\sc Sb}fg4 & 1 & $0^\circ,30^\circ,90^\circ,150^\circ,180^\circ$ &  0.4, 0.4 \\
 & {\sc Sc}fg4{\sc Im}fg4 & 4 & $0^\circ,30^\circ,90^\circ,150^\circ,180^\circ$ &  0.4, 0.4 \\
 & {\sc Sc}fg4{\sc Sd}fg4 & 2 & $0^\circ,30^\circ,90^\circ,150^\circ,180^\circ$ &  0.4, 0.4 \\
 & {\sc Sc}fg4{\sc Sc}fg4 & 1 & $0^\circ,30^\circ,90^\circ,150^\circ,180^\circ$ &  0.4, 0.4 \\
 & {\sc Sd}fg4{\sc Im}fg4 & 2 & $0^\circ,30^\circ,90^\circ,150^\circ,180^\circ$ &  0.4, 0.4 \\
 & {\sc Sd}fg4{\sc Sd}fg4 & 1 & $0^\circ,30^\circ,90^\circ,150^\circ,180^\circ$ &  0.4, 0.4 \\
 \hline
Gas--rich  & {\sc Sb}fg4{\sc Im}fg8 & 8 & $30^\circ,150^\circ$ &  0.4, 0.8 \\
Secondary & {\sc Sb}fg4{\sc Sd}fg8 & 4 & $30^\circ,150^\circ$ &  0.4, 0.8 \\
 & {\sc Sc}fg4{\sc Im}fg8& 4 & $30^\circ,150^\circ$ &  0.4, 0.8 \\
 & {\sc Sc}fg4{\sc Sd}fg8 & 2 & $30^\circ,150^\circ$ &  0.4, 0.8 \\
\hline
Gas--Rich  & {\sc Sb}fg8{\sc Im}fg8 & 8 & $30^\circ,150^\circ$ &  0.8, 0.8 \\
 & {\sc Sb}fg8{\sc Sd}fg8 & 4 & $30^\circ,150^\circ$ &  0.8, 0.8 \\
 & {\sc Sb}fg8{\sc Sc}fg8 & 2 & $30^\circ,150^\circ$ &  0.8, 0.8 \\
 & {\sc Sb}fg8{\sc Sb}fg8 & 1 & $30^\circ,150^\circ$ &  0.8, 0.8 \\
 & {\sc Sc}fg8{\sc Im}fg8 & 4 & $30^\circ,150^\circ$ &  0.8, 0.8 \\
 & {\sc Sc}fg8{\sc Sd}fg8 & 2 & $30^\circ,150^\circ$ &  0.8, 0.8 \\
 & {\sc Sc}fg8{\sc Sc}fg8 & 1 & $30^\circ,150^\circ$ &  0.8, 0.8 \\
 & {\sc Sd}fg8{\sc Im}fg8 & 2 & $30^\circ,150^\circ$ &  0.8, 0.8 \\
 & {\sc Sd}fg8{\sc Sd}fg8 & 1 & $30^\circ,150^\circ$ &  0.8, 0.8 \\ 
\hline
\hline
\end{tabular}
\label{tab:minor_sims}
\end{center}
\end{table*}

\begin{figure}
\plotone{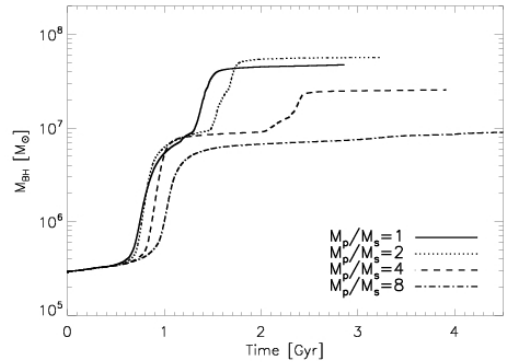}
\caption{SMBH growth histories for four different interactions with mass ratios of: $M_p/M_s \approx$ 8 (dash--double--dot), 4 (dash--dot), 2 (dash), and 1 (major merger; solid).  The orbital inclination for each was $i=30^\circ$ and the gas fractions of both initial disks 40\% ($f_g=0.4$).  The major mergers clearly grow much more massive SMBHs than do the minor ($M_{p}M_{s}\sim$4--8) mergers.}
\label{fig:minor_grow_massfrac}
\end{figure}

The interactions in the major mergers series, and in particular their
orbital parameters, are discussed in detail by 
\citet{robertson2006b,robertson2006a},
\citet{cox2006}, and
\citet{hopkins2007theory}; we refer the reader to these papers for
a detailed discussion.  Each galaxy typically 
contains $6\times 10^5$ halo particles (0.2\lengthunits softening), 
$2\times 10^4$ bulge particles (0.1\lengthunits softening), and 
$4\times 10^4$ stellar disk particles (0.1\lengthunits softening).  The two progenitor galaxies
are placed on a zero--energy parabolic orbit, as motived by cosmological N--body
simulations \citep[e.g.,][]{benson2005,khochfar2006}.  We include a
range of gas fractions, from relatively gas poor ($f_g = 0.05$) to pure gas
disks ($f_g = 1$), covering $\approx 2.5$ dex in total mass -- excluded the scaled 
disks from \citet{robertson2006a}, which are intended to be representative of high redshift systems.  
We have furthermore varied the resolution by up to a factor of 128 as many particles to confirm that our 
results are robust.

The mass ratio series (summarized in Table~\ref{tab:minor_sims})
includes all permutations of the progenitor disk models in
Table~\ref{tab:models}, which results in primary to secondary mass
ratios ranging from $M_p/M_s \approx 8$ to major mergers ($M_p/M_s =
1$).  Both initial disks are dynamically stable over a Hubble time,
and have disk mass fractions of $m_d = 0.05$.  The mass resolution of the halo component was
$\sim 4\times 10^6 M_{\odot}$ per particle (0.15\lengthunits
softening), while the mass resolution of the baryonic component (stars
and gas) was $\sim 3.5\times 10^5 M_{\odot}$ per particle
(0.05\lengthunits softening) -- the particle counts are summarized in Table~\ref{tab:minor_sims}.
To simplify the interpretation of the
results, none of these models initially had bulges.  As before, the two
disks are initialized on zero--energy parabolic orbits, with an impact
parameter of $\sim h_{D,p}$, where $h_{D,p}$ is the scale length of
the primary disk.  We also include five different orbital
inclinations, as in \citet{younger2007}: $i = 0^\circ$ (prograde
coplanar), $30^\circ$, $90^\circ$ (polar), $150^\circ$, and
$180^\circ$ (retrograde coplanar).  We consider three different
choices for the gas fractions of the initial disks: (1) a standard
gas--rich series \citep[as in][]{robertson2006a} where both initial
disks have $f_g=0.4$ over all mass ratios and inclinations, (2) a
series with more gas--rich lower--mass ({\sc Sd} and {\sc Im})
secondaries \citep[as suggested by observations;
see][]{roberts1994,bell2000,schombert2001,geha2006} with $f_{g,p} =
0.4$ and $f_{g,s} = 0.8$ for a prograde ($i=30^\circ$) and retrogade
($i=150^\circ$) interaction, and (3) a high gas fraction series with
$f_g=0.8$ for both initial disks \citep[which may be representative of
higher redshift progenitors; see][]{erb2006a,forster2006}, over all mass ratios
and again for a prograde and retrograde interaction.

\begin{figure*}
\plottwo{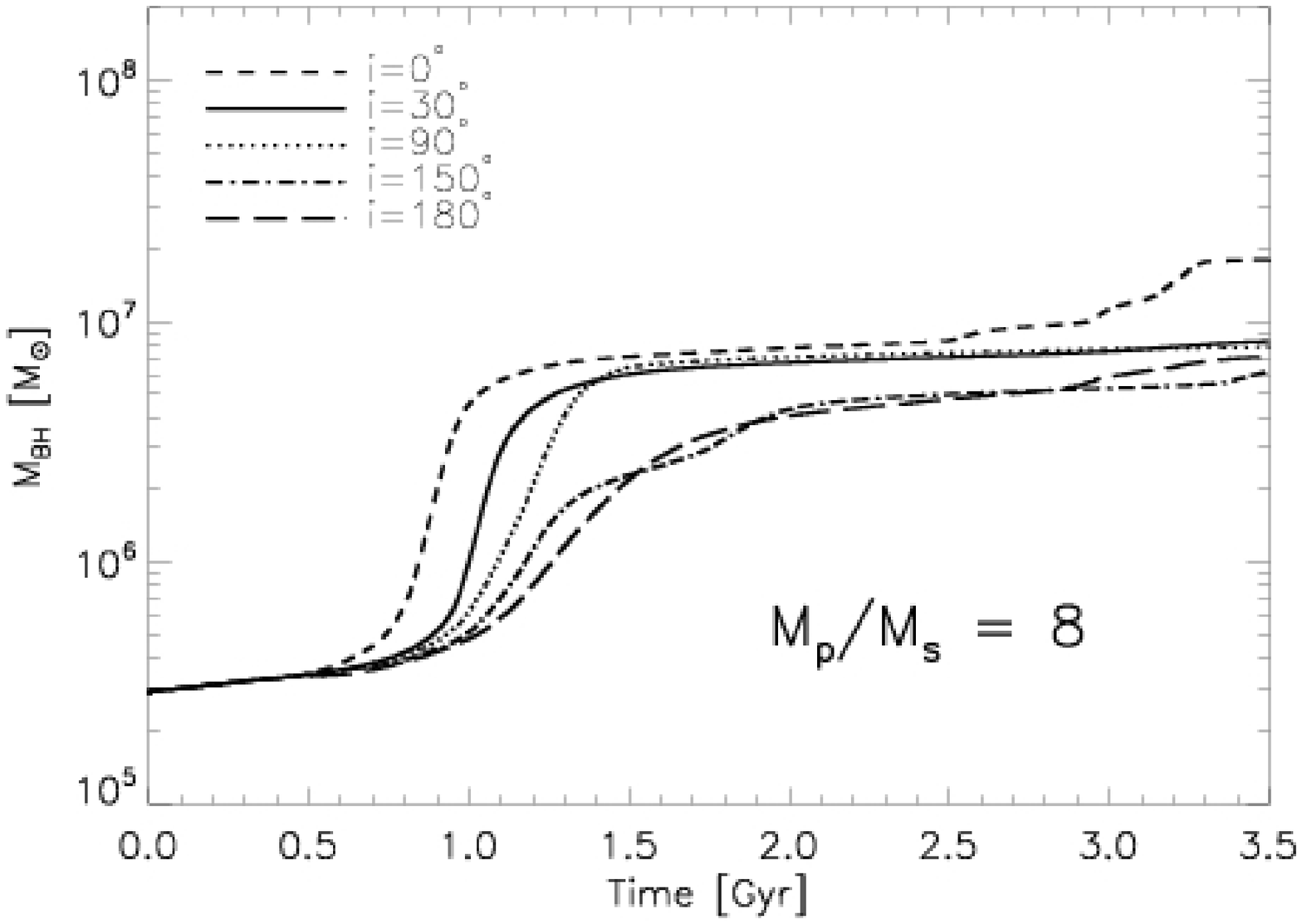}{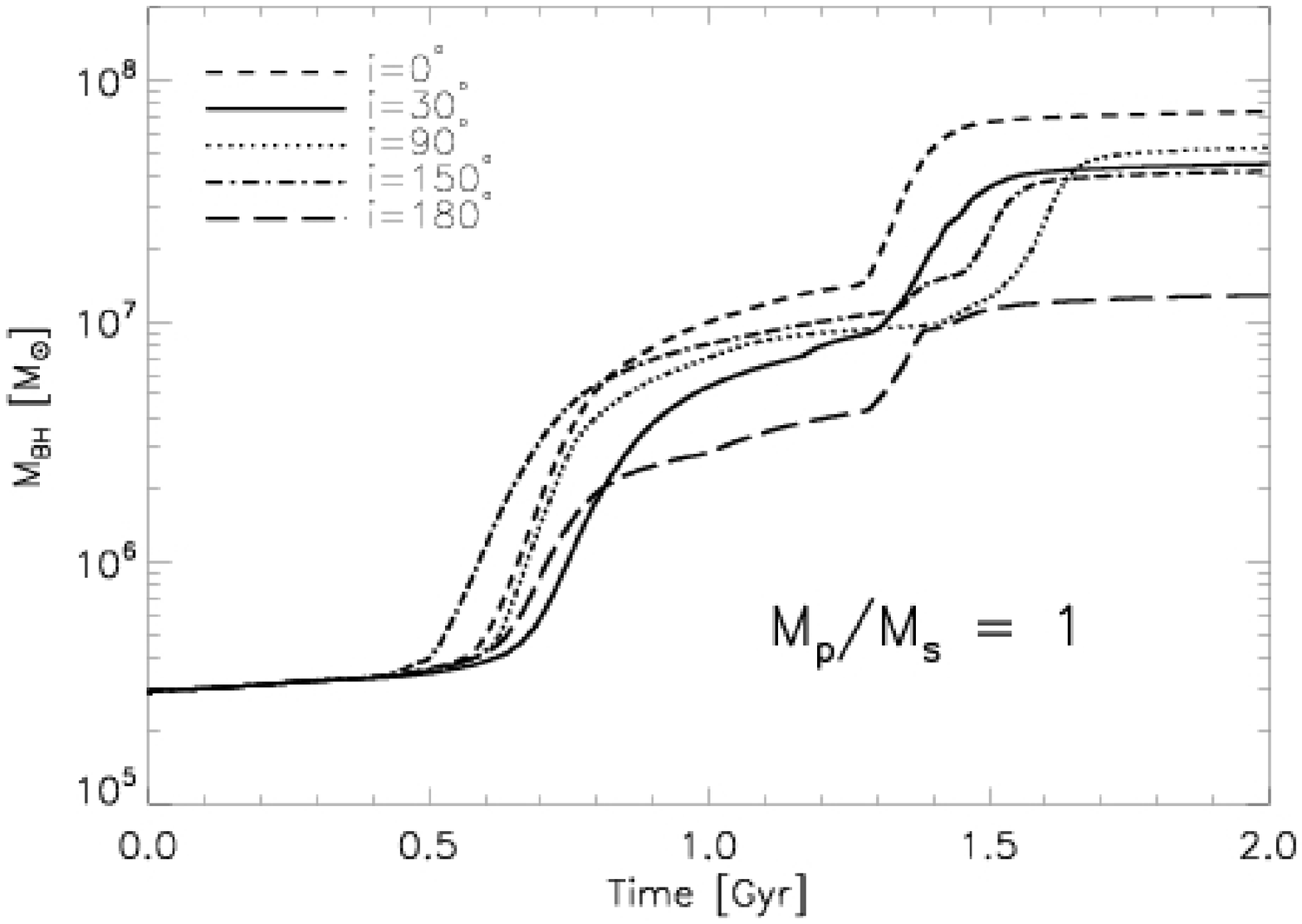}
\caption{SMBH growth histories for two different mass ratio interactions -- a major merger $M_p/M_s = 1$ (right panel) and minor merger $M_p/M_s \approx 8$ (left panel).-- for 40\% gas initial disks ( $f_g=0.4$) over  five different inclinations: $i=0^\circ$ (coplanar prograde; dash), $30^\circ$ (solid), $30^\circ$ (polar; dot), $150^\circ$ (dash--dot), $180^\circ$ (coplanar retrograde; dash--double--dot).  With the exception of exactly coplanar interactions, which are somewhat pathological, the final SMBH mass is not particularly sensitive to the orbital geometry.}
\label{fig:minor_grow_inc}
\end{figure*}

The process by which SMBHs grow and self--regulate in major mergers is
described in detail by several authors \citep[see
e.g.,][]{SDH2005,hopkins2005c,hopkins2005d,hopkins2006a}.  Broadly,
tidal torques during close passages between disks induce bar
instabilities in the stellar distribution.  These stellar bars then drain the gas within a 
critical radius of its angular momentum 
\citep{hopkins2008h}
causing it to flow towards the central regions,  which in turn fuels a nuclear starburst \citep[see
also][]{barnes1991,mihos1994,mihos1996,barnes1996} and grows the
SMBH nearly at an exponential rate.  Eventually, when the
accretion rate reaches a critical value, feedback terminates
the gas inflow and drives a galaxy scale ``superwind" that shuts down
both star formation and SMBH growth.  At the same time, the
interaction dynamically heats the stellar component, transforming it
into a pressure-supported spheroid: a ``red and dead"
bulge--dominated galaxy \citep[see also][]{barnes1992a,barnes1992b,
hernquist1992,hernquist1993a}.

When the mass ratio is sufficiently high ($M_p/M_s\gsim 4$; a ``minor
merger"), the interaction does not entirely destroy the initial disk 
\citep[for a detailed discussion for disk survival see][]{hopkins2008h},
but instead moves a smaller fraction of the stellar mass into a
central bulge component \citep[for a more detailed discussion of the
kinematics of disk heating by minor mergers, see
e.g.][]{quinn1986,quinn1993,walker1996,velazquez1999,kazantzidis2007}.
And, in general the nuclear starburst and SMBH accretion do not
exhaust the available gas supply, leaving a substantial fraction in
the disk component of the remnant \citep[see also][for images and a
detailed analysis]{cox2008}.

These clear differences motivate a look at the growth history of SMBHs
in our simulations, and their sensitivity to the parameters of the
interaction.  A detailed investigation into the efficiency with which
mergers grow black holes, and their luminous lifetimes \citep[see
e.g.,][]{hopkins2005b,hopkins2005d,hopkins2006a} for different mass
ratios is a topic worthy of more detailed study, which we postpone to
future work.  However, as a qualitative illustration, we show the
growth histories of SMBHs in mergers of different mass ratios and
orbital inclinations.

In Figure~\ref{fig:minor_grow_massfrac}, we see that at fixed
inclination and primary galaxy mass, major mergers ($M_p/M_s \sim
2-1$) grow SMBHs that are nearly an order of magnitude more massive
than those produced in minor mergers ($M_p/M_s \gsim 4$).  As shown by
\citet{hopkins2007theory} and in subsequent sections of this work, the
SMBH traces the binding energy of the bulge.  Therefore, the
efficiency of SMBH growth in interactions of differing mass ratio is
likely a consequence of the efficiency of dynamical heating and bulge
formation during those encounters.  

Qualitatively, in a minor merger, the first passage induces a bar, gas
inflow \citep{hernquist1989a,hernquistmihos1995} and -- if it is an interpenetrating 
encounter -- violently relaxing approximately $M_s$ of the disk stars, which 
together form a bulge \citep[see][]{hopkins2008h}\footnote{This suggests that 
as the impact parameter increases, the resultant bulge will tend more towards a pseudobulge.  
In fact, since no galaxy is truly isolated, the unstable disks outlined in \S~\ref{sec:unstable_disks} can 
be thought of as the limiting case of low mass satellites and halo substructure seeding the 
bar instabilities via flyby encounters.  We have run a limited experiment, varying the impact 
parameter for a small subset of interactions and find evidence that the structural properties of the 
resultant bulges do tend towards pseudobulges for very large impact parameters (including 
the normalization offset in $M_{BH}-\sigma$; see \S~\ref{sec:projected_correlations} and 
\ref{sec:discussion_msigma}).  However, a systematic investigation into the dependence of 
the structural properties of the bulge on the orbital parameters of the interaction 
is outside the scope of this paper, and thus we defer it to future work.}.  
It also tends to put the secondary on a mostly radial orbit, and strip a significant fraction of its
mass \citep{bullock2005,abadi2006,stewart2008} prior to final coalescence.  A low
angular momentum final orbit, coupled with a central mass
concentration will tend to damp out further bar--driven gas inflows
\citep[see e.g.,][and discussion in
\S~\ref{sec:unstable_disks}]{shen2004,hozumi2005,bournaud2005,debattista2006}.
This, in combination with less dynamical heating owing to the lower
energetics of the interaction, suppresses further bulge growth, and
with it the final burst of SMBH activity seen in major mergers.

\begin{table*}
\begin{center}
\caption{Initial Parameters for the Unstable Disk Series}
\begin{tabular}{cccccc}
\hline
\hline
& Name &  Disk Mass Fraction & Initial Gas Fraction & Seed BH Mass & $q_{EOS}$ \\
& & $m_d$ & $f_{g}$ & $M_{BH,i}$/(\msolarunits) & \\
\hline
Standard & {\sc uSb}08 & 0.08 & 0.4, 0.6, 0.8 & $10^5$  & 1.0\\
& {\sc uSc}08 & 0.08 & 0.4, 0.6, 0.8 & $10^5$  & 1.0 \\
& {\sc uSd}08 & 0.08 & 0.4, 0.6, 0.8 & $10^5$  & 1.0 \\
& {\sc uSb}10 & 0.10 & 0.4, 0.6, 0.8 & $10^5$  & 1.0 \\
& {\sc uSc}10 & 0.10 & 0.4, 0.6, 0.8 & $10^5$  & 1.0 \\
& {\sc uSd}10 & 0.10 & 0.4, 0.6, 0.8 & $10^5$  & 1.0 \\
\hline
BH Seed Mass & {\sc usSb}10 & 0.10 & 0.4 & $10^6$ & 1.0 \\
& {\sc usSc}10 & 0.10 & 0.4 & $10^6$ & 1.0 \\
& {\sc usSd}10 & 0.10 & 0.4 & $10^4$ & 1.0 \\
\hline
EOS & {\sc usSb}10 & 0.10 & 0.4 & $10^5$ & 0.5 \\
& {\sc usSc}10 & 0.10 & 0.4 & $10^5$ & 0.5 \\
& {\sc usSd}10 & 0.10 & 0.4 & $10^5$ & 0.5 \\
\hline
\end{tabular}
\label{tab:unstable_sims}
\end{center}
\end{table*}

\begin{table*}
\begin{center}
\caption{Resolution of the Unstable Disk Series}
\begin{tabular}{cccccc}
\hline
\hline
Name &  $N_{halo}$ & $N_{baryons}$ \\
\hline
{\sc uSb}08 & $3.2\times 10^5$ & $3.2\times 10^5$ \\
{\sc uSc}08 & $1.7\times 10^5$ & $1.7\times 10^5$ \\
{\sc uSd}08 & $7.9\times 10^4$ & $7.8\times 10^4$ \\
{\sc uSb}10 & $3.2\times 10^5$ & $4.0\times 10^5$ \\
{\sc uSc}10 & $1.7\times 10^5$ & $2.1\times 10^5$ \\
{\sc uSd}10 & $7.7\times 10^4$ & $9.8\times 10^4$ \\
\hline
\end{tabular}
\label{tab:unstable_resolution}
\end{center}
\end{table*}

In Figure~\ref{fig:minor_grow_inc}, we examine the effect of orbital
inclination -- prograde coplanar ($i=0^\circ$) to retrograde coplanar
($i=180^\circ$) -- on both minor (left panel) and major (right panel)
mergers.  In general, compared to retrograde interactions, prograde
encounters produce stronger tidal responses in the disk that induce
bars and drive the gas inflows that fuel bulge formation and SMBH
growth \citep[see e.g.,][]{cox2008}.  We see this effect in our
simulations: while the effect is more dramatic for major mergers, for
the limiting case of a coplanar interaction, retrograde orbits result
in a significantly less massive SMBHs than do prograde.  However, when
the spins of the two galaxies are not aligned, the resulting SMBH is
largely independent of orbital inclination.

These trends with mass fraction and orbital inclination qualitatively
motivate the aim of this work: given a particular encounter, what
mechanism sets the final SMBH mass?

\subsection{Disk Instabilities}
\label{sec:unstable_disks}
The simulations in the unstable disk series evolve variants of the same disk models in isolation, the parameters of which are summarized in Table~\ref{tab:unstable_sims}.  The mass resolution and softening lengths of the particles were kept the same as with the mass ratio series, and the particle counts are summarized in Table~\ref{tab:unstable_resolution}.  The disk mass fraction is increased while holding the total mass fixed, which seeds global instabilities by increasing the self--gravity of the disk relative to its kinetic energy from rotation \citep[see][]{ostriker1974,efstathiou1982}.  For three initial disk models ({\sc Sb, Sc,} and {\sc Sd}), we consider two disk mass fractions ($m_d = 0.08$ and $0.10$), and three gas fractions ($f_g = 0.4,0.6,$ and $0.8$).  For a subset of these simulations, we consider three different seed black hole masses -- $M_{BH,i} = 10^4, 10^5$, and $10^6$ \msolarunits -- and two different equations of state -- $q_{EOS} = 0.5$, 1.0.

Stellar bars are found generically in numerical simulations of isolated disk galaxies \citep{hohl1971}, and represent a global instability in which rotating disks swing--amplify spiral density wave perturbations \citep{toomre1981,binney1987}.  They have been shown empirically to develop when the self--gravity of the disk is comparable to its kinetic energy of rotation \citep{ostriker1974,efstathiou1982}.  Most disk galaxies in the local universe have bars \citep[e.g.,][]{eskridge2000,menendez2007}, and -- despite conflicting observational results -- it seems clear that a substantial fraction of high redshift $z\sim 1$ disks also show bar features \citep[e.g.,][]{jogee2004,sheth2008}.  These bars have been shown both theoretically \citep{athanassoula1992,athanassoula2000} and observationally to drive gas inflows and nuclear starbursts \citep{jogee1999,jogee2002,jogee2005,petitpas2002,petitpas2003}.  As a result, they represent a potentially promising mechanism for fueling the growth of nuclear SMBHs, which has at times been invoked in semi--analytic models \citep[e.g.,][]{bower2006} as their dominant mode of growth.

Since the onset of the instability is a collisionless process \citep[see][]{toomre1981,binney1987}, it develops first in the stellar distribution.  The stellar bar then drains the gas within a critical radius of its angular momentum \citep{barnes1996,hopkins2008h}, driving it inwards fueling a nuclear starburst; a process analogous to merger--driven gas inflows in which the bar is induced by the time--evolution of the tidal field during close passages \citep[see][]{mihos1994}.  This nuclear starburst concentrates some of the mass of the initial disk at its center, forming a pseudobulge \citep[see][for a review]{kormendy2004} and eventually leading to the destruction of the bar over a relatively short timescale as the central mass concentration damps out the spiral density waves \citep[$\sim 1-2$ Gyr;][]{shen2004,hozumi2005,bournaud2005,debattista2006}.  Because there is no interpenetration by a satellite, there is correspondingly no violent relaxation of the stellar disk and as a result the pseudobulge is almost entirely dominated by stars formed during the simulation.

\begin{figure}
\plotone{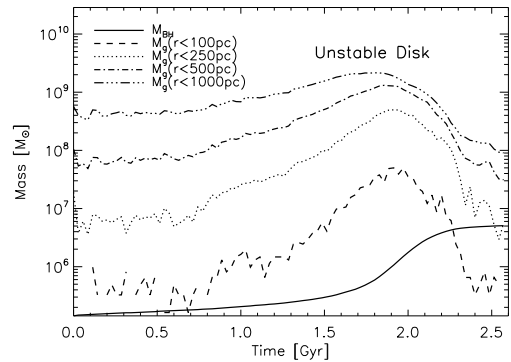}
\caption{Illustration of the bar--driven gas inflows and resulting feedback driven outflows in a representative unstable disk simulation: {\sc uSb08} with baryon fraction $m_d=0.08$ and initial gas fraction $f_g=0.4$.  Shown is the SMBH mass (solid line) and total gas mass ($M_g$) within 100 (dashed line), 250 (dotted line), 500 (dash--dotted line), and 1000 (dash--triple--dotted line) parsecs.  The stellar bar efficiently concentrates a significant gas reservoir -- several hundreds $\times M_{BH}$ -- within a few hundred parsecs of the SMBH which are subsequently expelled via feedback driven outflows.}
\label{fig:bars_gasmass}
\end{figure}

\begin{figure}
\plotone{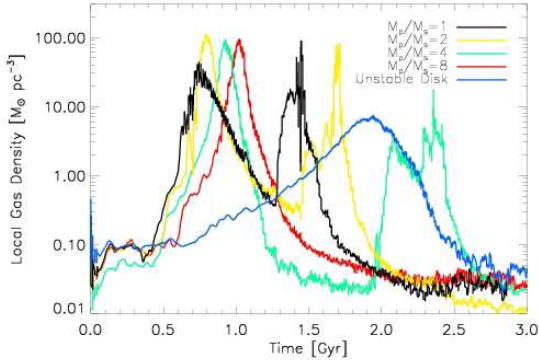}
\caption{Local gas density -- the mean density over one SPH smoothing length of 64 particles -- near one of the SMBHs in four merger simulations of varying mass ratio -- $M_p/M_s = 1$ (black line), 2 (yellow line), 4 (green line), and 8 (red line), all with an orbital inclination of $i=30^\circ$ and initial gas fraction $f_g=0.4$ -- as compared to a representative unstable disk simulation (same as in Figure~\ref{fig:bars_gasmass}; blue line).  The local gas densities, which are used to calculate the instantaneous accretion rate of the SMBH, are comparable in mergers and unstable disks.}
\label{fig:bars_localrho}
\end{figure}

Though the feedback model employed for the unstable disk series is identical in both implementation and parameter choice to that used for the major mergers and mass ratio series, in order to insure a fair comparison it is important that the growth of the SMBH be self--regulated by feedback rather than limited by gas depletion in the central regions.  This will be the case if the stellar bars, like those induced during close passages during major and minor mergers, drive gas inflows that create a central gas reservoir many times more massive than the SMBH and a substantial density enhancement in its immediate vicinity, which will result in Eddington--limited growth until the local gas supply is then expelled by a thermal feedback driven blast wave.    In Figure~\ref{fig:bars_gasmass} we show the gas inflow -- and feedback driven outflow -- as a function of time for a representative unstable disk simulation.  We find that the gas inflows driven by the stellar bar concentrates a significant gas supply in the central several hundreds of parsecs.  Furthermore, in Figure~\ref{fig:bars_localrho} we confirm that  the gas density enhancement in the immediate vicinity of the SMBH -- which is used to calculate the instantaneous accretion rate -- is comparable to those produced in merger--driven inflows.  Finally, we find that the accretion rate -- i.e., AGN light curve -- decays rapidly with a power--law slope similar to that expected from feedback driven outflows \citep[e.g.,][]{hopkinshernquist2006}.  As a result,  we confirm that SMBH growth in our unstable disk simulations is regulated by feedback rather than gas depletion, and thus -- if we assume that accretion and feedback in mergers and unstable disks are similar -- our feedback model is appropriate to these systems.

\subsection{Systematic Uncertainties}

\citet{robertson2006a} and \citet{hopkins2007theory} both comment on
the sensitivity of the predicted SMBH scalings arising from major
mergers to choices in our models: in particular the choice of the
initial seed mass $M_{BH,i}$ and the EOS parameter $q_{EOS}$, which
unlike $f_g$ do not have a clear observationally preferred, physically
motivated value.  These authors find that over a relatively wide range of
values for these parameters, the scaling relations with final SMBH are
well within the scatter both in the observed and simulated relations.
Therefore, while the overall growth history of an individual
simulation may change significantly with specific choices for these
parameters, the predicted scalings remain statistically unchanged.
This is consistent with our interpretation of the scaling of SMBH mass
with the binding energy of the bulge as the more fundamental, physical
relation; the parameters of the bulge component of the remnant are
relatively insensitive to these parameters.

Because the mass ratio series is consistent with the scalings
derived from major mergers, we are confident that they too are not
affected by these systematic uncertainties in our modeling.  However,
it is not clear {\it a priori} how varying $M_{BH,i}$ and $q_{EOS}$
will affect SMBHs grown in unstable disks.  This is because the SMBH
does not grow by more than one or two orders of magnitude and the
stochastic nature of its accretion may make it more sensitive to the
details of the growth history than SMBHs grown via major or even minor
mergers, which accrete several orders of magnitude more at comparable
total baryonic mass (see, e.g. Figure~\ref{fig:minor_grow_massfrac}).

\begin{figure}
\plotone{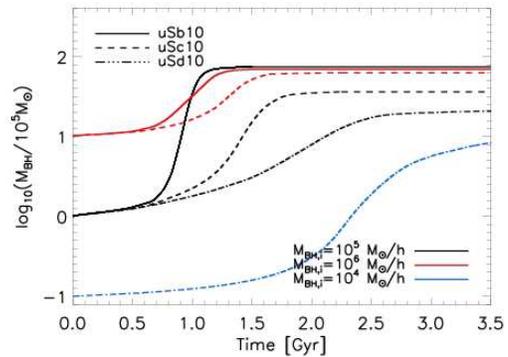}
\caption{SMBH growth histories for three different disk models -- {\sc Sb} (solid line), {\sc Sc} (dashed line), and {\sc Sd} (dot--dot--dashed line) -- and three different SMBH seed masses -- $M_{BH,i} = 10^4$ (blue), $10^5$ (black), and $10^6$ (red) $h^{-1} M_\odot$.  We find that, despite differences in the SMBH growth histories, there is no systematic trend between the final SMBH mass and $M_{BH,i}$, and that the maximum deviation for an order of magnitude difference in the seed mass is only $\sim 0.3$ dex.}
\label{fig:bars_mbh_grow_sbh}
\end{figure}

\begin{figure}
\plotone{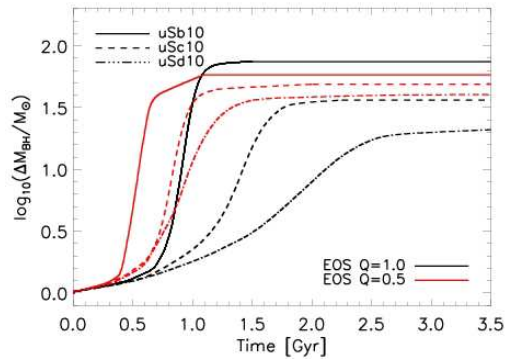}
\caption{SMBH growth histories for three different disk models -- {\sc Sb} (solid line), {\sc Sc} (dashed line), and {\sc Sd} (dot--dot--dashed line) -- and three different SMBH seed masses -- $q_{EOS} = 0.5$ (red) and 1.0 (black).  We find -- as with the initial seed SMBH mass -- despite differences in the growth histories there is no systematic trend between the final SMBH mass and the stiffness of the EOS.}
\label{fig:bars_mbh_grow_eos}
\end{figure}

In Figure~\ref{fig:bars_mbh_grow_sbh} we show the growth history of
SMBHs in three different disk models with $f_g=0.4$ for different
seed masses: {\sc uSb10} ($M_{BH, i} = 10^5, 10^6 M_{\odot}/h$), {\sc
uSc10} ($M_{BH, i} = 10^5, 10^6 M_{\odot}/h$), and {\sc uSd10}
($M_{BH, i} = 10^4, 10^5 M_{\odot}/h$).  We find that, as with the gas
supply, while the detailed evolution differs substantially, the
final SMBH masses are largely consistent to within the scatter in the
simulated and observed relations.  The largest difference ($\sim 0.3$ dex)
is for the lowest mass disk model, for which the SMBH accretes the
least mass and is therefore more sensitive to certain choices of
parameters for our modeling.

In Figure~\ref{fig:bars_mbh_grow_eos}, we show the SMBH growth history
for different effective equation of state parameters $q_{EOS}=0.5$ and
1.0 for three different disk models with $f_g=0.4$.  As $q_{EOS}$
approaches unity, it increases the dynamical stability of the gas
against \citet{toomre1964} instabilities
\citep[see][]{robertson2004,robertson2006a}.  This effectively delays
the onset of the bar--driven inflow in disk models with $q_{EOS} =
1.0$ relative to those with a softer EOS $q_{EOS} = 0.5$.  However,
despite these differences in the growth history of the SMBHs, their
final masses are roughly consistent with one another to within the
scatter in the observed and simulated relations.  The largest
differences, as with different seed masses, are for the lowest mass
disk model ($\sim 0.3$ dex).  More important, there are no
systematic trends with $q_{EOS}$: the final SMBH mass in uSb10 is
somewhat lower, while those in uSc10 and Sd10 are slightly higher.

\subsection{Remnant Properties}
\label{sec:methods_remnant}

\begin{figure*}
\plottwo{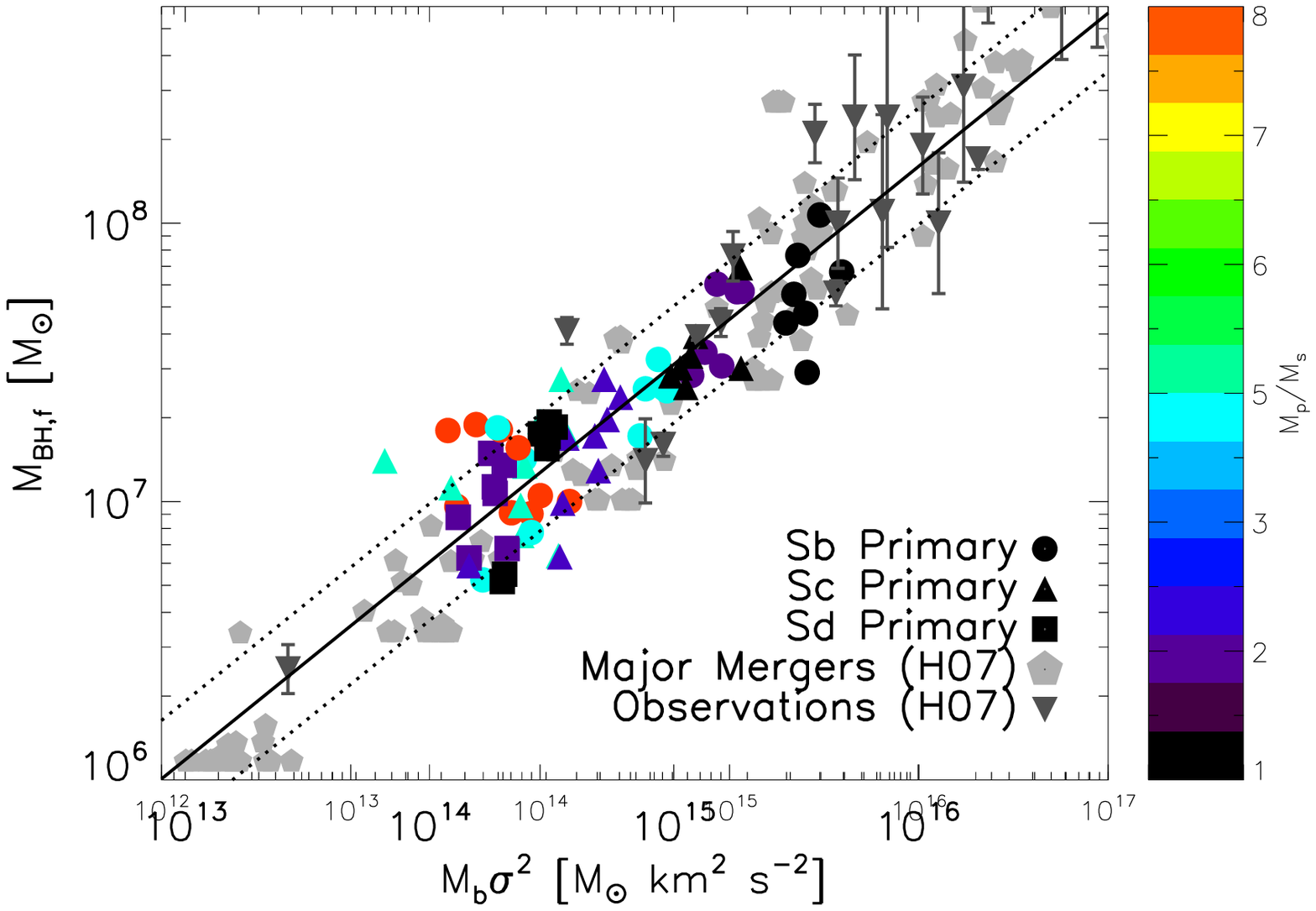}{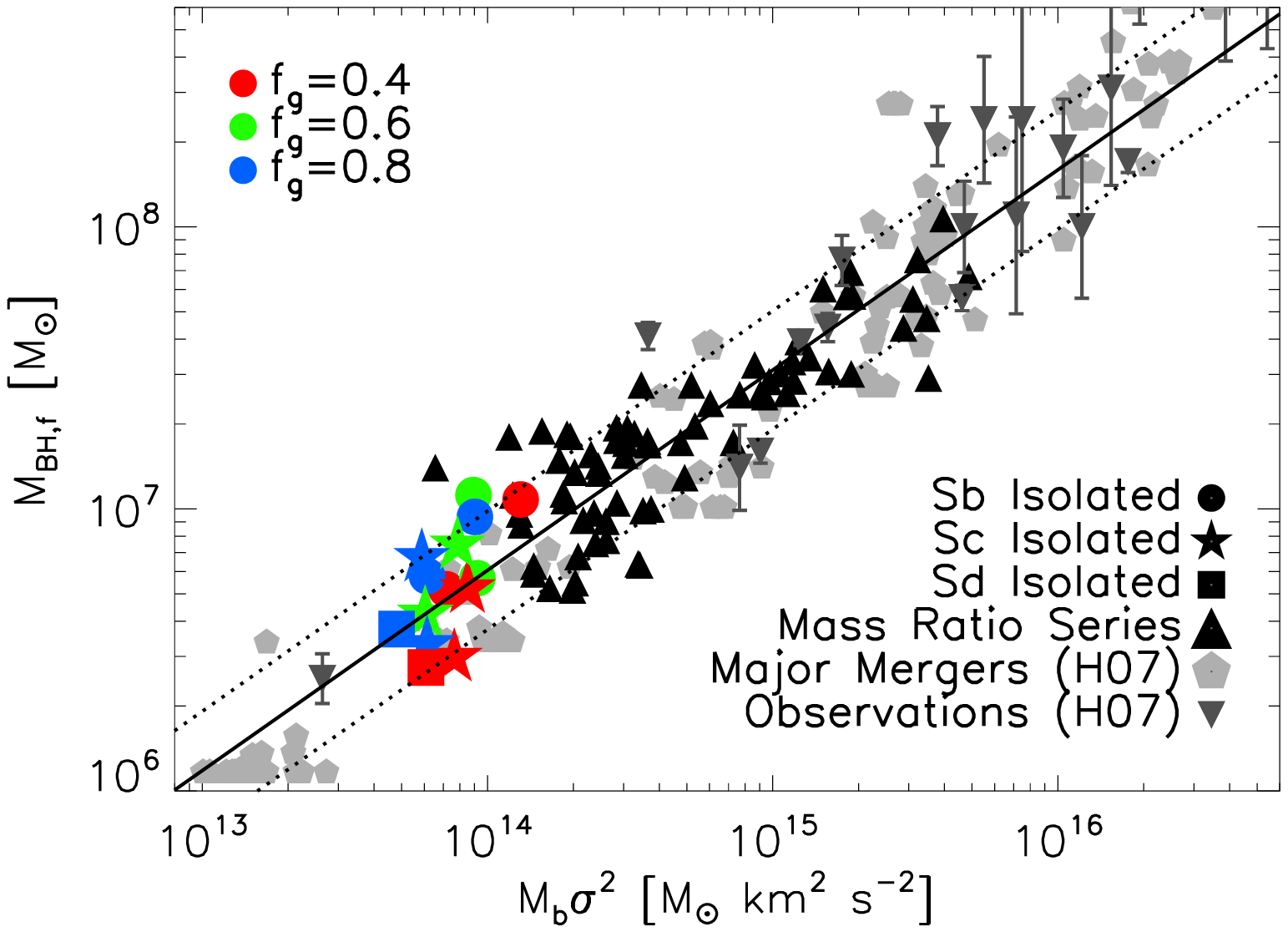}
\caption{Left: Binding energy correlation -- which is statistically equivalent to the BHFP \citep[H07;][]{hopkins2007theory} -- for the major and mass ratio series (solid line, scatter indicated by dotted lines).  The grey hexagons are the major mergers series and the grey inverted triangles are observations, both from H07.  The mass ratio series -- including all gas fractions -- are colored according to the mass ratio (see color bar to the right of the figure) with the primary galaxy model indicated according to: {\sc Sb} (filled circle), {\sc Sc} (filled triangles), and {\sc Sd} (filled square).  Right: The binding energy correlation for mergers (see previous; the mass ratio series as filled triangles) and unstable disks (same labeling as primary galaxy in previous) for three different gas fractions: $f_g = 0.4$ (red), 0.6 (green), and 0.8 (blue).  We find that SMBHs grown via major mergers, minor mergers, and disk instabilities all lie along the same BHFP and binding energy correlations.}
\label{fig:bhfp_all}
\end{figure*}

\begin{figure}
\plotone{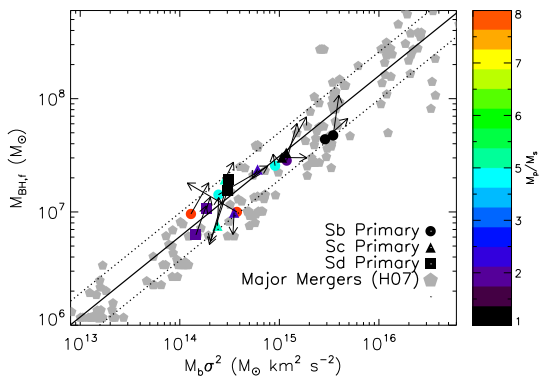}
\caption{Influence of gas fraction on the location of SMBHs on the binding energy correlation for the mass ratio series. The labeling is the same as the left panel of Figure~\ref{fig:bhfp_all}.  Shown are the mass ratio series simulations with initial gas fraction $f_g=0.4$, and the arrows indicate the location of identical interactions with more gas rich progenitors ($f_g=0.8$).  We find that increasing the gas content of the progenitors drives these systems along these relations -- owing to the structural properties of bulges formed during more gas--rich interactions \citep[see e.g.,][]{robertson2006c,robertson2006b} -- but not systematically away from them.  This reflects the different structural properties of bulges formed from more gas--rich disks, leaving the fundamental character of the BHFP and binding energy correlations unchanged.}
\label{fig:bind_gas_move}
\end{figure}

The structural and kinematic properties of the remnant were extracted
after the system reached a state of approximate dynamical equilibrium;
for the major and mass ratio series, this was typically $\sim
1h^{-1}$ Gyr after the final coalescence of the two SMBHs, while for
the unstable disks this was typically $\sim 0.5 h^{-1}$ Gyr after the
peak of the SMBH accretion rate.  If the remnant was disk--dominated,
we then fit a combination exponential and \citet{sersic1968} profile
to the projected stellar mass density (i.e., ``surface brightness'')
profile along 64 lines of sight\footnote{The detailed formation and
growth of stellar bulges via minor mergers is certainly an interesting
problem in its own right, and the simulations presented here are ideal
for just such an analysis.  However, for this work we restrict
ourselves to simple profile fits and bulge masses for both the
purposes of clarity and to best approximate observational estimates of
bulge masses, and postpone a detailed analysis of the bulge structure
and kinematics to a future paper (Cox et al. 2008, in preparation)}, from
which we estimated the median bulge mass ($M_b$) and effective radius
($R_e$).  If the remnant was bulge--dominated (B/T$\gsim 0.8$), we took $M_b$ to be the
total stellar mass, and $R_e$ the half--mass radius.  The inner
velocity dispersion ($\sigma$) was measured within $R_e$, again taking
the median along 64 different lines of sight.  This procedure was identical 
between the mergers and unstable disks.  Furthermore, \citet{hopkins2008h} show
that surface brightness fitting of the kind employed in this work recovers on average the 
bulge--to--disk ratios inferred from a kinematic decomposition of the stellar particles.

\section{Results}
\label{sec:results}

\subsection{The Black Hole Fundamental Plane}
\label{sec:bhfp_results}

\begin{figure*}
\plottwo{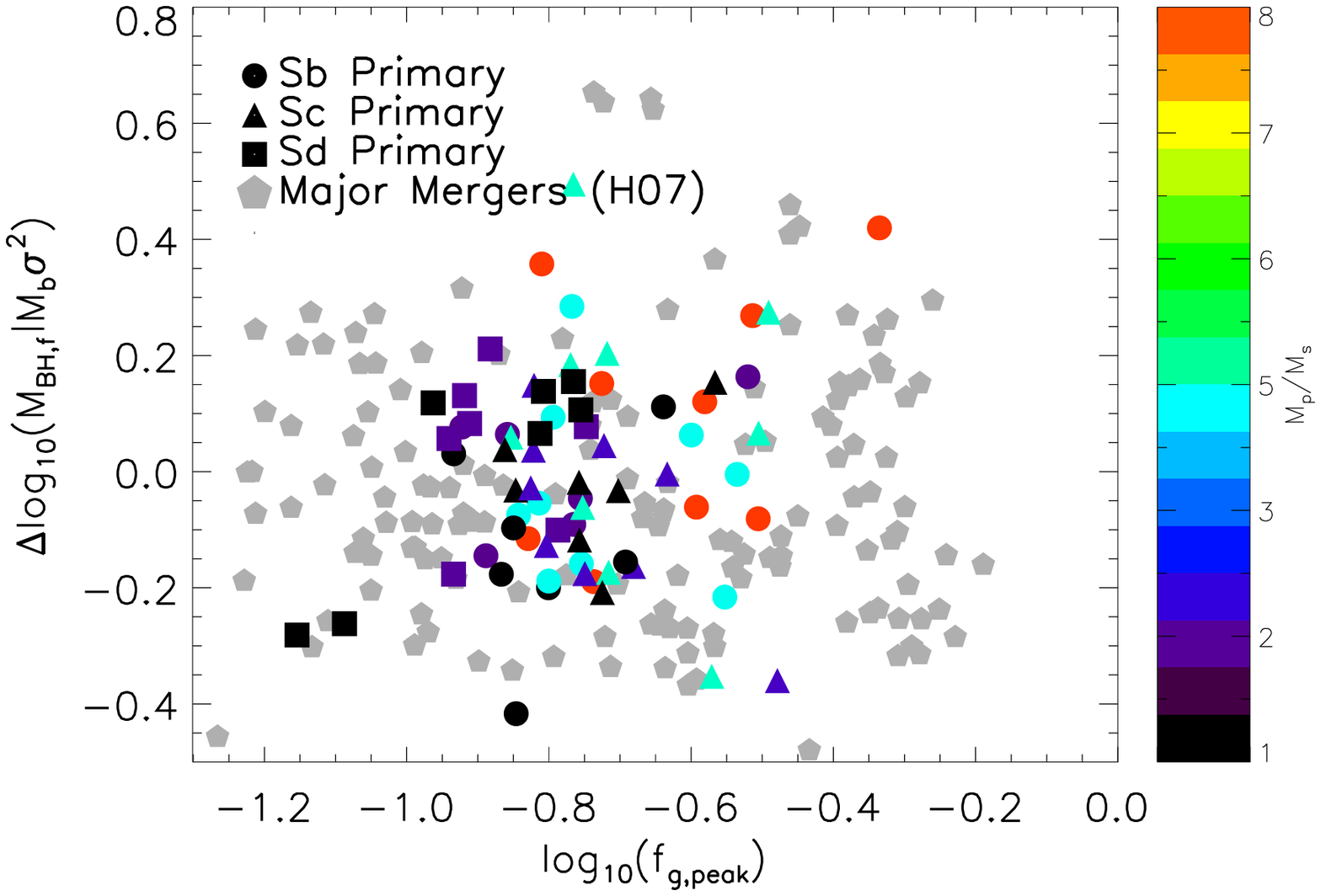}{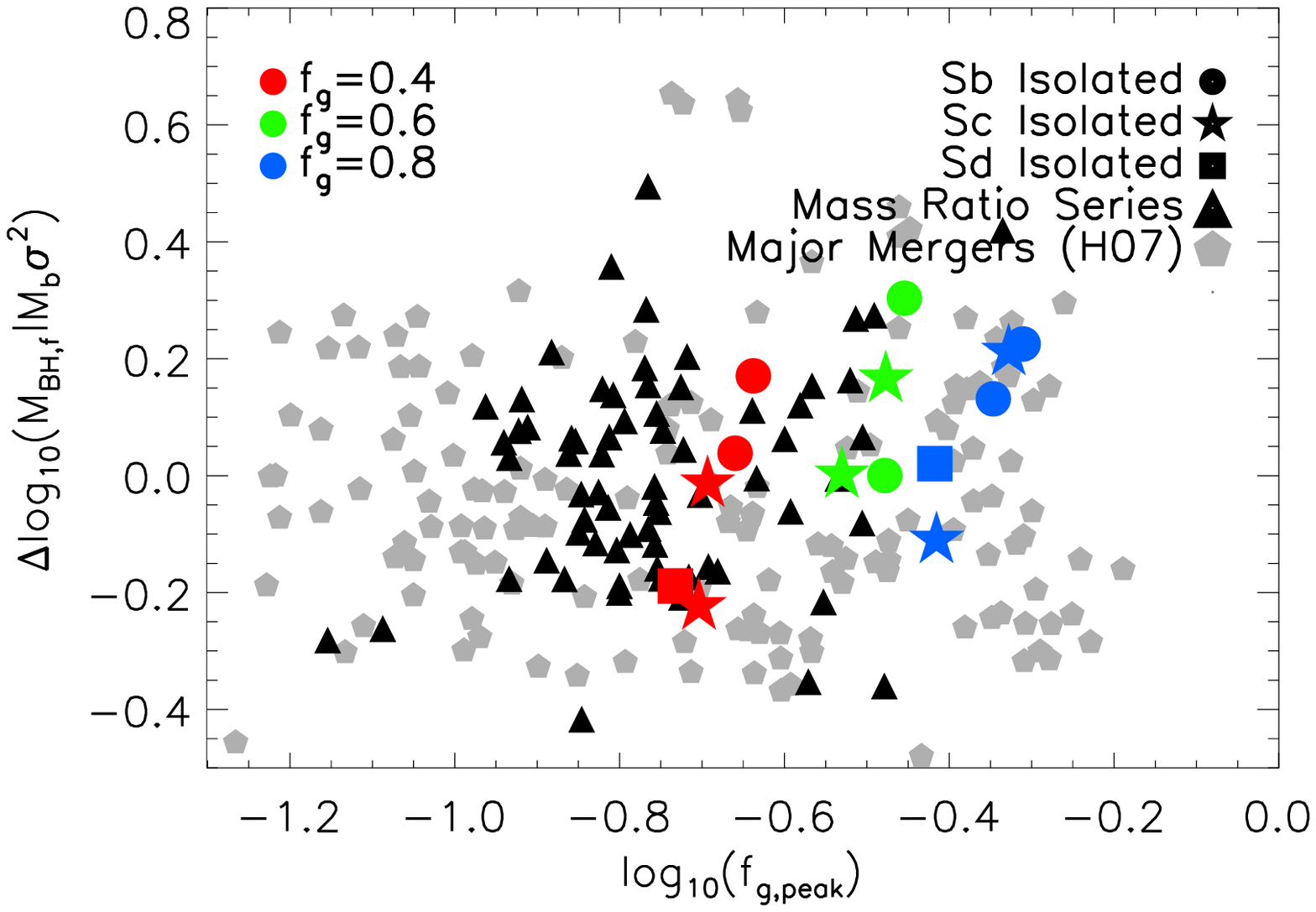}
\caption{Correlation between the SMBH mass residual at fixed binding energy and the gas fraction at the peak of the starburst ($f_{\rm g,peak}$) for mergers (left) and unstable disks (right), with the same labeling as Figure~\ref{fig:bhfp_all}.  We find that these residuals are uncorrelated in all three sets of simulations, indicating that the final SMBH mass is not determined by the available fuel supply.}
\label{fig:gassupply_all}
\end{figure*}

\begin{figure*}
\plottwo{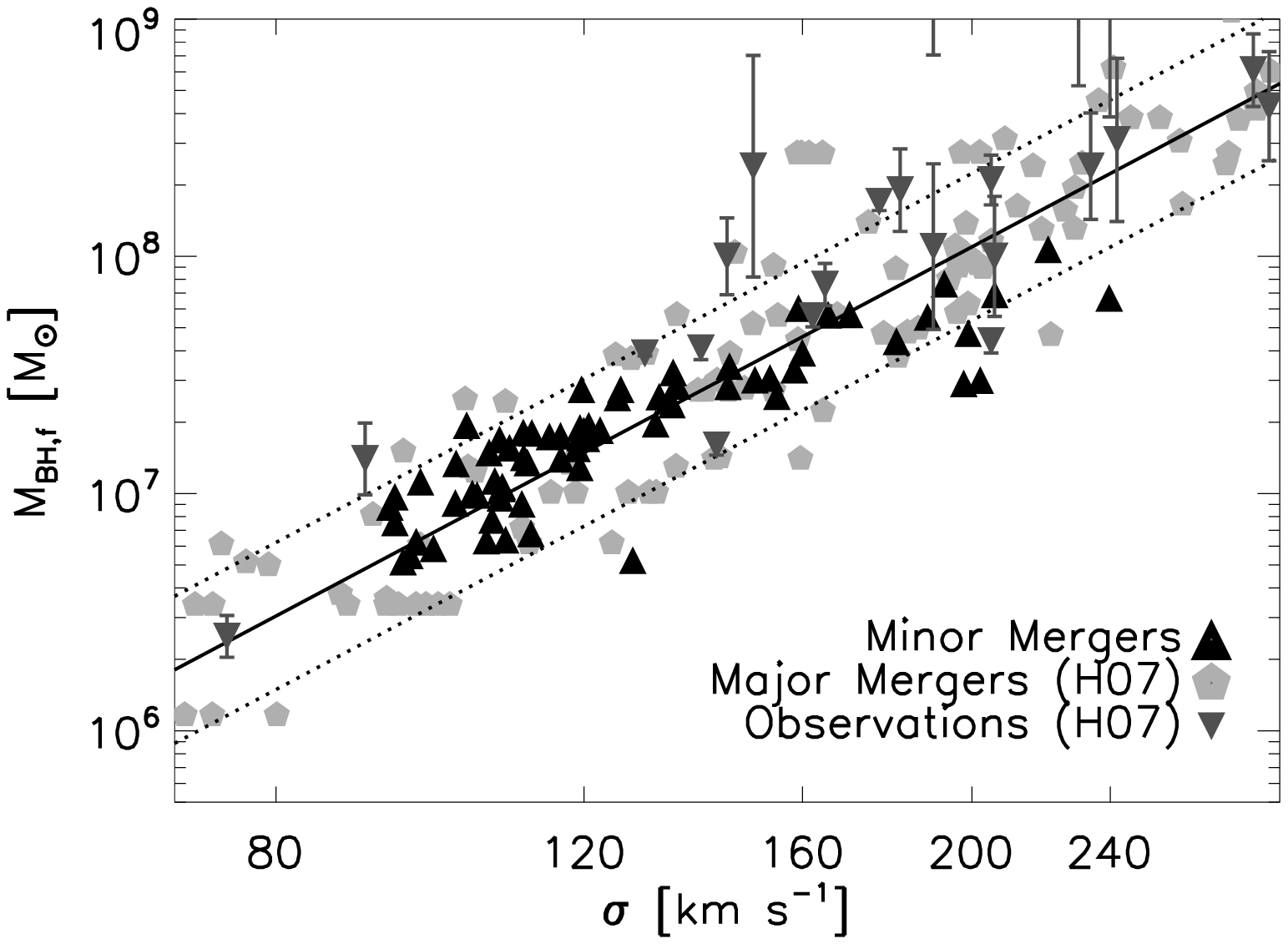}{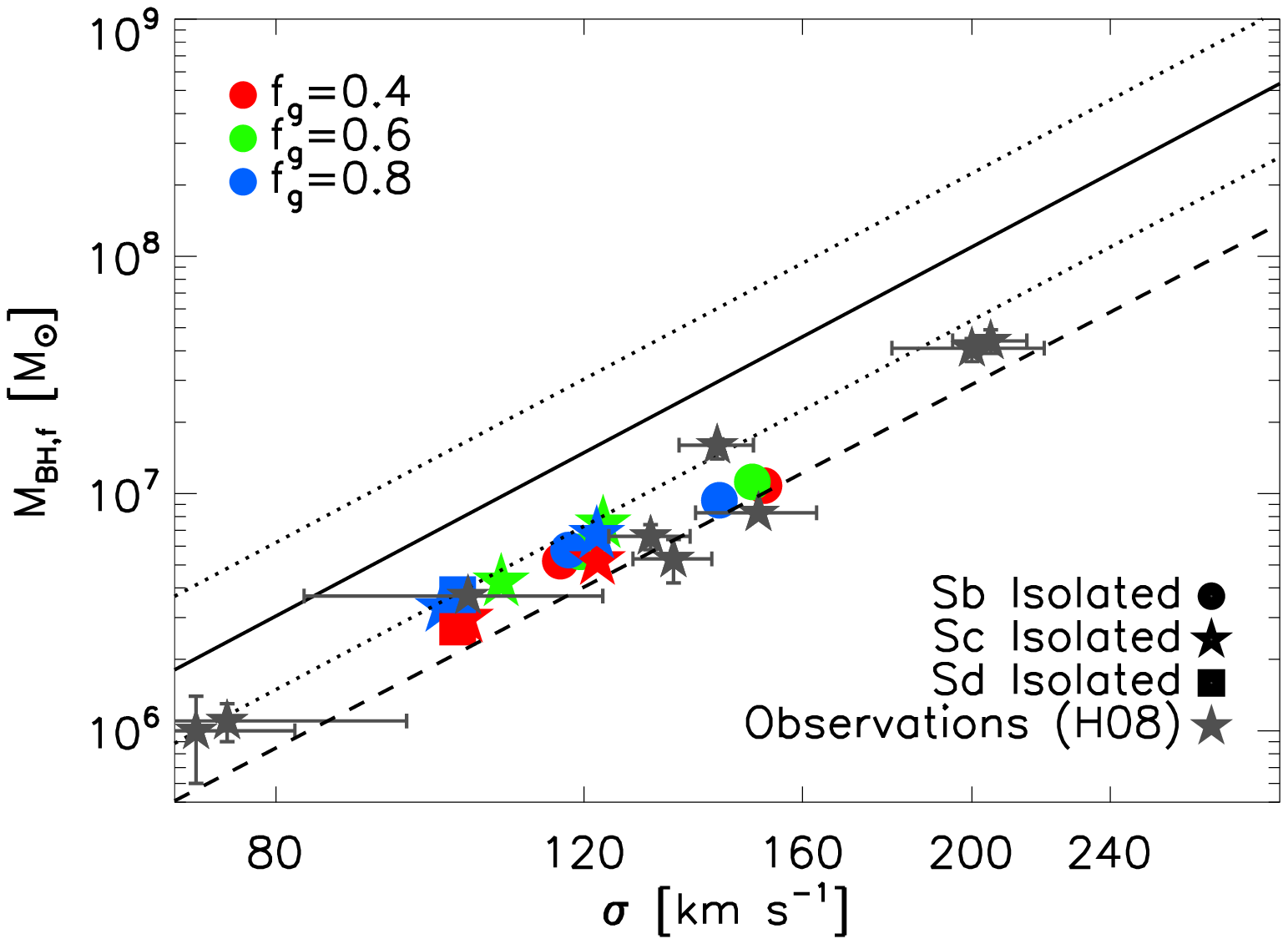}
\caption{$M_{BH}-\sigma$ relation for mergers (left) and unstable disks (right), with the same labeling as the right hand panel of Figure~\ref{fig:bhfp_all}.  The black solid and dashed lines indicate the best--fit relations and scatter from H07, and we include observations of SMBHs hosted by pseudobulges -- which are likely grown via disk instabilities \citep[see][]{kormendy2004} -- from \citet[H08; grey stars indicate data, dashed line indicates best--fit relation;][]{hu2008}.}
\label{fig:msigma_all}
\end{figure*}

\citet{hopkins2007theory} find that there exist correlations in the
residuals of fits such as the $M_{BH}-\sigma$ and
\citet{magorrian1998} relations; for example, at fixed $\sigma$ the
residual SMBH mass scales approximately as $\sim M_{b}^{0.72}$, and at
fixed $M_{b}$ the residuals scale as $\sim \sigma^{1.40}$.  By
marginalizing over two parameters -- $M_{b}$ and either $\sigma$ or
$R_e$ -- they found a best--fit BHFP \citep[analogous to that observed for
elliptical galaxies;][]{dressler1987,djorgovski1987}
which minimized these residual correlations with the form $M_{BH}\sim M_b^{0.72}\sigma^{1.4}$, both in simulations \citep{hopkins2007theory} and for observed systems
\citep{barway2007,hopkins2007obs,aller2007}.  This is statistically indistinguishable 
from a correlation with the bulge binding energy proxy $M_b\sigma^2$ ($M_{BH}\sim (M_b\sigma^2)^{0.7}$), and can be interpreted as reflecting the nature of
feedback regulated SMBH growth: accretion accelerates until feedback
is sufficient to unbind the local gas supply, abruptly terminating the
inflow and cutting off further growth.  Therefore it is more
``fundamental" than its various projections, such as the
$M_{BH}-\sigma$ and \citet{magorrian1998} relations -- a point we
discuss in more detail in \S~\ref{sec:discussion_fundamental}.

To systematically test this hypothesis, we examine the binding energy
correlation -- which is statistically equivalent to the BHFP -- in
three different modes of SMBH fueling: major mergers, minor mergers,
and unstable disks (see \S~\ref{sec:sims} for details).
Figure~\ref{fig:bhfp_all} shows the binding energy correlation
for major mergers
(grey hexagons), the mass ratio series (left: colored points; right: black
triangles), and unstable disks (right: colored points), along with the
observations listed in \citet{hopkins2007obs}.  We find that over a
range of baryonic masses, gas fractions, and orbital parameters they
all lie along the same relation to within the scatter, and that all
reproduce the observed correlation.

\subsection{The Role of the Fuel Supply}
\label{sec:bhfp_fuel}

Inasmuch as their growth is terminated at a critical accretion
rate, the final masses of SMBHs should not be determined by the
available fuel supply so long as the gas reservoir is much more massive
than the SMBH.  While it is the case that the final SMBH mass is
strongly correlated with the total gas mass of the system, this
reflects the structural properties of bulges in gas--rich merger
remnants: owing the effects of dissipation, a more gas rich progenitor
will lead to a more compact bulge at fixed total mass, and
consequently a deeper central potential $\phi_c$ and velocity
dispersion $\sigma$ \citep[see
e.g.,][]{robertson2006b,robertson2006a,hopkins2007theory}.  As shown
by \citet{hopkins2007theory} for major mergers, at fixed potential the
gas fraction has no effect on SMBH growth.

One method of illustrating this is presented in
Figure~\ref{fig:bind_gas_move}, which shows the BHFP and binding
energy correlations for the mass ratio series (colored points) as
compared to those derived from simulations of major mergers (grey
hexagons; solid line with the dotted line indicating the scatter).
Here we show that increasing the initial gas fraction from $f_g=0.4$
to $0.8$ for the identical interaction will drive the remnant along
the BHFP and binding energy correlations, but not systematically away
from them.  Therefore, we find that the typically more massive SMBHs
resulting from simulations of more gas--rich encounters is related to
the structural properties of the resulting remnant -- which is more
compact at fixed $M_b$
\citep{hopkins2008a,hopkins2008b,hopkins2008c,hopkins2008d,
hopkins2008e},
and thus has larger binding energy -- than it
is to the larger available fuel supply, leaving the nature of these
correlations unchanged.

In Figure~\ref{fig:gassupply_all}, we show the residual correlation of SMBH mass at fixed binding energy with the gas fraction at the peak of the starburst ($f_{\rm g,peak}$) for both mergers (left) and unstable disks (right).  We find no correlation between this residual and $f_{\rm g,peak}$.  This demonstrates again that to the extent that SMBH growth is self--regulated, its final mass is not correlated with the available fuel supply but rather with the binding energy of its host bulge.

\subsection{The Projected Correlations}
\label{sec:projected_correlations}

\begin{figure}
\plotone{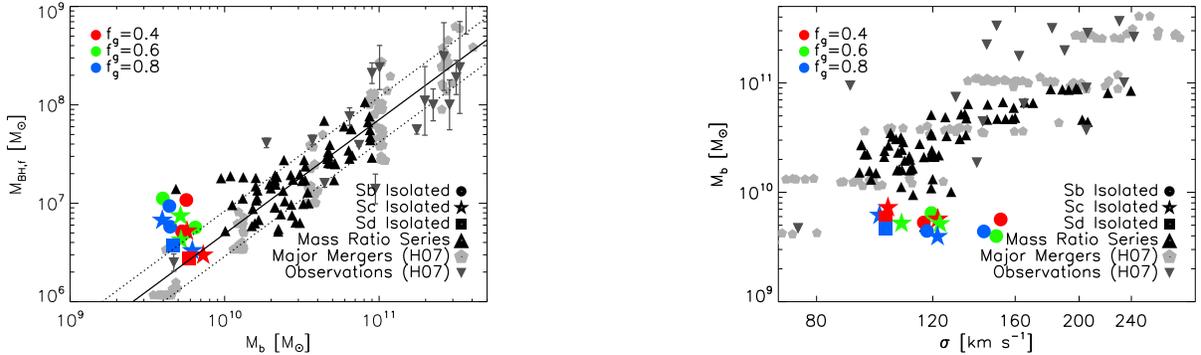}
\caption{\citet{magorrian1998} relation for mergers and unstable disks, with the same labeling as Figure~\ref{fig:msigma_all}.}
\label{fig:magorrian_all}
\end{figure}

While the BHFP and binding energy scalings with SMBH mass are consistent between mergers and unstable disks, individual  projections of these relations are not consistent.  In Figures~\ref{fig:msigma_all} and \ref{fig:magorrian_all} we show the predicted $M_{BH}-\sigma$ and \citet{magorrian1998} scalings from the unstable disk simulations, as compared to both the major merger and mass ratio series.  One of the more striking features of these results is the normalization offset in the $M_{BH}-\sigma$ relation, in which SMBHs grown via disk instabilities are less massive at fixed $\sigma$ than those grown in mergers but maintain a similar slope.  Or conversely, the bulges grown via disk instabilities are more compact at fixed SMBH and bulge mass.  

\begin{figure}
\plotone{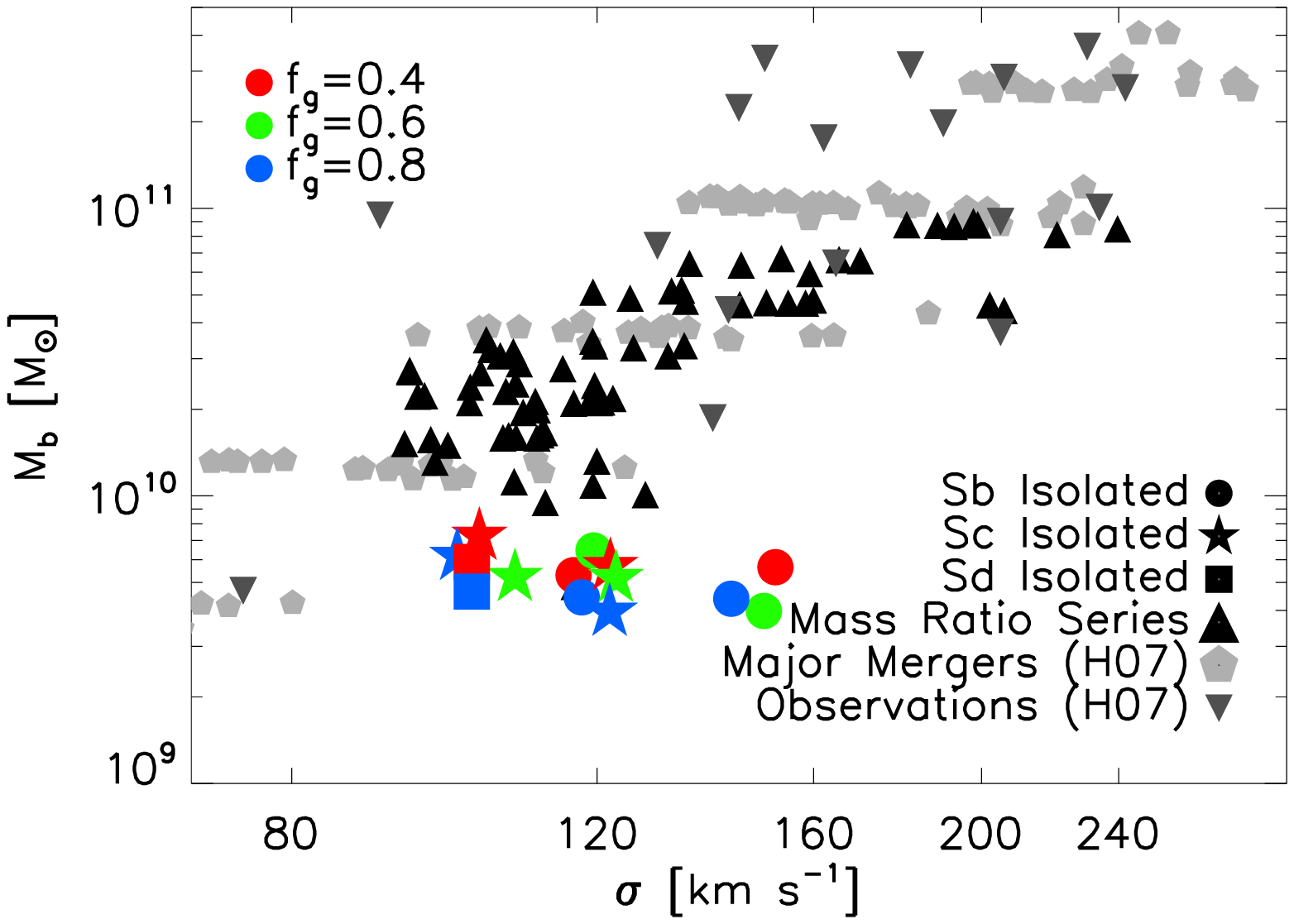}
\caption{\citet{faber1976} relation for bulges grown via mergers and disk instabilities, with the same labeling as the right hand panel of Figure~\ref{fig:bhfp_all}.  Bulges produced via major and minor mergers (``classical bulges") lie along the same correlation, while those produced via secular processes \citep[``pseudobulges";][]{kormendy2004} follow a different relation.  This demonstrates that these two types of bulges are structurally different; plausible, given their likely different formation mechanisms.}
\label{fig:faberjackson_all}
\end{figure}

This difference was noted by \citet{hu2008} and \citet{graham2008} as
a normalization offset between the $M_{BH}-\sigma$ relations in
``classical" bulges, which are likely the result of mergers, and
``pseudobulges" or bulges in barred systems which were likely grown
via secular processes \citep{kormendykennicutt2004}.  Our
simulations agree quite well with these observational results.
\citet{hu2008} interprets this result as possible evidence for less
efficient fueling of SMBHs via bar instabilities, creating a less
massive SMBH at fixed $\sigma$.  However, we find that this reflects
structural differences between bulges produced in merger simulations
versus those produced via secular processes in the unstable disks:
bulges and pseudobulges\footnote{The exact location of the
pseudobulges produced in unstable disks on the \citet{faber1976} relation is somewhat sensitive to the
initial conditions.  However, their manifestly different structural
properties relative to classical bulges produced via mergers is a
robust result.} lie along different \citet{faber1976} relations (see
Figure~\ref{fig:faberjackson_all}).  These systems all lie along the
same BHFP and binding energy correlations, suggesting that the lower
mass SMBHs found in pseudobulges at fixed $\sigma$ reflect their lower
overall binding energy owing to structural differences rather than the
efficiency of fueling.  These structural differences are driven largely by the 
different formation mechanism of pseudobulges, which retain a great deal more 
rotation than their classical analogues -- the LOS velocity dispersion will include 
rotational motions.  In addition, because all the mergers 
presented in this work coalesce, the structural differences can further be understood
via simple energy arguments: in a merger, conservation of kinetic
energy requires that the effective radius of the bulge increase -- and
therefore the velocity dispersion must decrease -- when the system has
fully merged \citep[e.g.,][]{hernquist1993}.  

\section{Discussion}
\label{sec:discussion}

\subsection{The ``Fundamental" Character of the BHFP}
\label{sec:discussion_fundamental}

Among local systems, there are a relatively large number of tight
correlations between SMBHs and the properties of their host galaxies
\citep[for reviews, see
e.g.,][]{novak2006,hopkins2007theory,graham2008}.  The question of
which of these is more ``fundamental" has two components: (1) which is
a better predictor of SMBH mass in systems for which detailed
measurements are not feasible, and (2) which better reflects the
physical mechanism driving the co-evolution of SMBHs and bulges.  The
answer to (1) is largely determined by observational uncertainty and
sample selection, and is outside the scope and limits of this
investigation.  However, (2) is directly addressed by our simulations,
and highlights the physical mechanisms causing the self--regulated
growth of SMBHs.

\citet{graham2008} argues that the residual--residual correlations
that initially motivated the observed and simulated BHFPs are a
consequence of sample selection, which includes bulges produced via both
interactions and secular processes.  In particular, the author
notes that when barred systems are excluded, these correlations are no
longer significant.  This is entirely consistent with expectations
from our modeling.  Owing to the relatively small dynamic range probed
by classical bulges with robust SMBH mass measurements, we would not
necessarily expect the residual correlations to appear significant
without the inclusion of lower mass systems, which are more likely to
be barred.  More important, the correlation between residuals
will be most apparent in systems with different structural properties,
like classical bulges and massive ellipticals as compared to
pseudobulges and barred systems.

\begin{figure*}
\plottwo{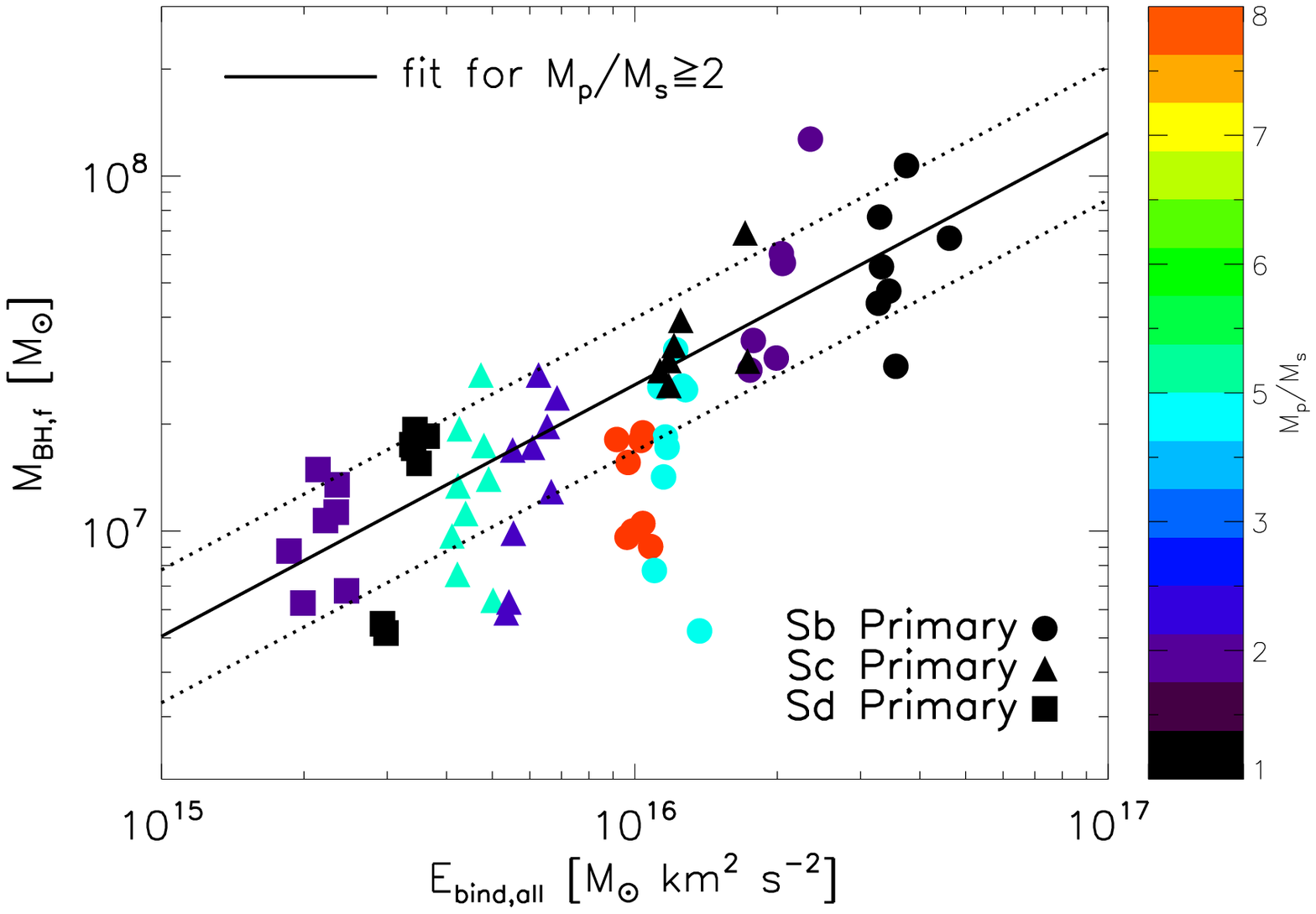}{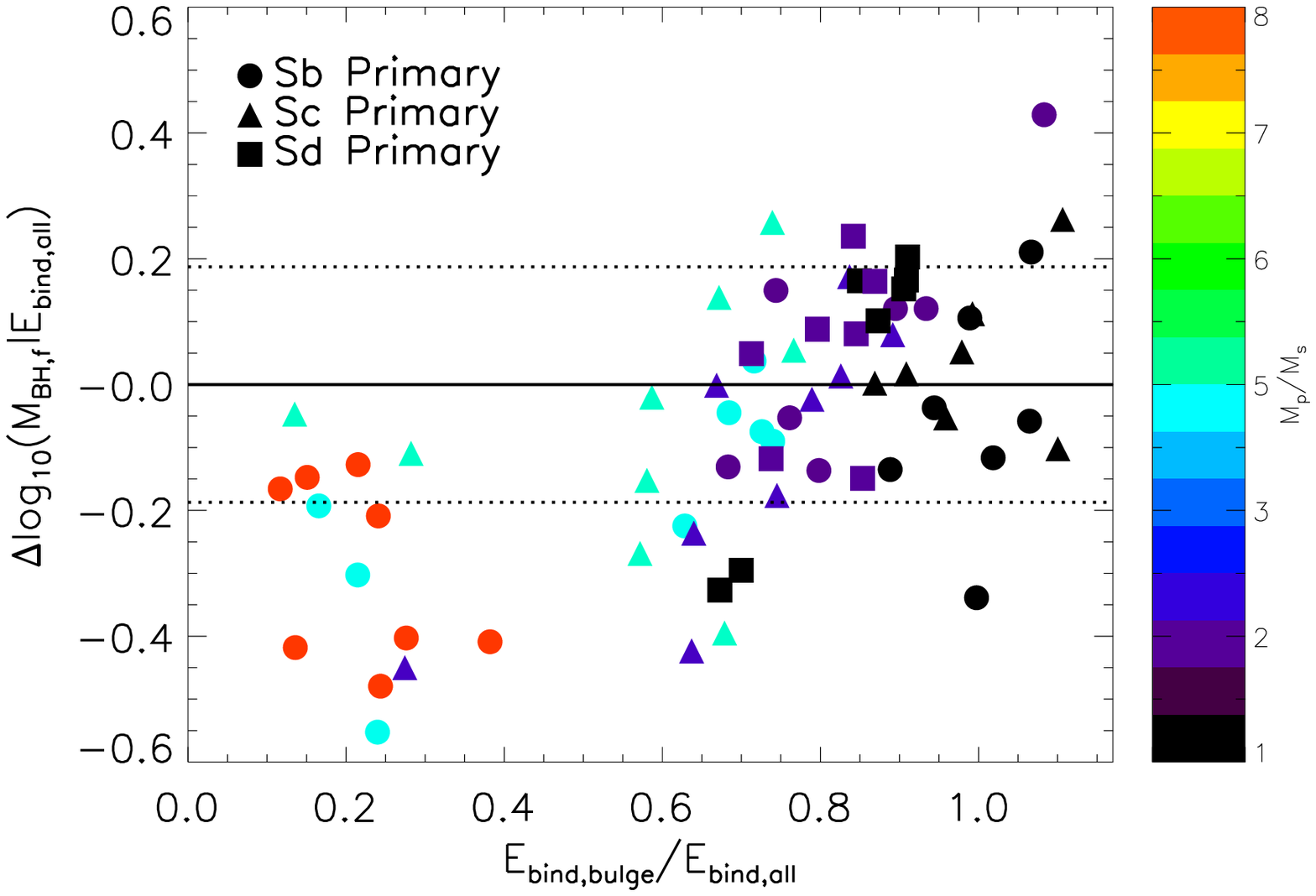}
\caption{Left: Correlation between SMBH mass and the binding energy of all the stellar particles $E_{\rm a,bind}$, for the mass ratio series.  Labeling is the same as the left hand panel of Figure~\ref{fig:bhfp_all}. The solid line represents a fit to the relation for major mergers ($M_p/M_s \lsim 2$), which are bulge--dominated, with the r.m.s. scatter indicated by dotted lines.  Right: The residual SMBH mass at fixed $E_{\rm all,bind}$, as compared to $E_{\rm b,bind}/E_{all,bind}$ using the relation derived for major mergers (see left), where $E_{\rm bulge,bind}\approx 10.1M_{b}\sigma^2$ is the approximate total binding energy assuming a spherical \citet{hernquist1990} profile.  We find a significant residual correlation as $E_{\rm b,bind}/E_{all,bind}$ approaches unity, suggesting that the bulge binding energy is a better predictor of final SMBH mass than the total stellar binding energy in systems that retain a significant disk component (minor mergers; $M_p/M_s \gsim 4$); i.e., the SMBH ``sees" the bulge more so than the overall stellar distribution.}
\label{fig:bulge_all_bind}
\end{figure*}

Moreover, our results suggest that while the results of
\citet{graham2008} directly address (1), they do not provide
convincing evidence against the BHFP and binding energy argument as
the solution to (2).  In Figure~\ref{fig:bhfp_all}, we show that
simulations of three different fueling mechanisms -- major mergers,
minor mergers, and disk instabilities -- all reproduce the observed
BHFP and binding energy correlations.  This suggests that in all three
cases, the mass of the SMBH is set by the critical accretion rate at
which feedback terminates the gas inflow, and halts further growth of
the SMBH.  At the same time, SMBHs grown in unstable disks versus
interactions lie along different projected correlations in
Figures~\ref{fig:msigma_all} and \ref{fig:magorrian_all}.  In this
way, the BHFP and binding energy correlations are more fundamental in
that they reflect the feedback regulated growth of SMBHs across bulges
with very different structural properties.

\subsection{Implications for SMBH Fueling Models}
\label{sec:discussion_fueling}

Our modeling demonstrates that in feedback--regulated models of SMBH
growth, while the BHFP and binding energy correlations are universal,
their projections -- $M_{BH}-\sigma$, \citet{magorrian1998}, etc. --
are sensitive to the structural properties of the spheroidal component
of the host galaxy.  For the particular case of $M_{BH}-\sigma$, this
results in a normalization (see \S~\ref{sec:projected_correlations}
and Figure~\ref{fig:msigma_all}) offset between classical bulges --
which were likely formed during mergers -- and pseudobulges and those
found in barred systems -- which are likely produced via secular
processes.  Given that our particular implementation offers a good
approximation to the effects of feedback more broadly (see
\S~\ref{sec:methods_feedback}), this is a generic prediction of models
in which SMBH growth is self--regulated via feedback.  

Consequently,
that both the BHFP \citep{aller2007,hopkins2007obs} and an apparent
offset in $M_{BH}-\sigma$ between classical and secular bulges
\citep{hu2008,graham2008} are observed provides evidence for feedback
regulated models more generally
\citep{silk1998,king2003,king2005,murray2005,sazonov2005,robertson2006a,thacker2006,hopkins2007theory}.
It also presents a challenge to alternative models in which the projected
scalings are relatively insensitive to the structural properties of
the spheroid, including: stellar capture by the accretion disk
\citep{zhao2002,MiraldaEscude2005}, regulation via a viscous
star--forming accretion disk \citep{burkert2001}, adiabatic SMBH
growth \citep{macmillan2002}, and direct gas collapse
\citep{adams2001,adams2003}.

\subsection{Implications for Evolution in the $M_{BH}-\sigma$ Relation}
\label{sec:discussion_msigma}

The existence of these two complementary $M_{BH}-\sigma$ relations
also has potential consequences for measurements of the evolution of
this correlation.  Studies of AGN at intermediate redshift have led to
somewhat ambiguous conclusions: some show evolution in $M_{BH}-\sigma$
\citep{treu2004,treu2007,walter2004,woo2006,woo2008}, while others do not
\citep{shields2003}.  Measurements of SMBH masses at intermediate
redshift are done using reverberation mapping techniques
\citep{blandford1982,peterson1993}, which use spectral variability to
infer the size scale of the broad line region and -- assuming circular
orbits -- estimate the mass of the SMBH \citep[see
also][]{wandel1999,kaspi2000b}.  These virial relations are calibrated
using a relatively small sample of local, relatively low--luminosity
AGN
\citep{gebhardt2000b,ferrarese2001,onken2004,nelson2004}, many of 
which may reside in pseudobulge hosts and therefore calibrating them as classical 
bulges may not be appropriate.

However, at cosmological distances, AGN will be preferentially more
luminous, and therefore more likely to be fueled by mergers than disk
instabilities.  This could lead to inferred evolution in the
$M_{BH}-\sigma$ relation which is more related to selection effects
than true evolution in the correlation.  In fact, the shift of $\Delta
{\rm log}\sigma \approx -0.15$ observed by \citet{treu2004,treu2007}
and \citet{woo2006,woo2008} is entirely consistent simply with the difference
between the normalizations of the merger and disk instability
$M_{BH}-\sigma$ relations in Figure~\ref{fig:msigma_all}.

\subsection{Why the Bulge?}

Both observational and theoretical work has demonstrated that the SMBH
is correlated with the structural properties of its host galaxy.
However, these correlations are far more significant when applied to the
spheroidal component -- or bulge -- rather than the galaxy as a whole.
We illustrate this in Figure~\ref{fig:bulge_all_bind}, in which we
show the binding energy correlation including all the stellar
particles ($E_{\rm bind,all}$; left), and the residual SMBH mass at
fixed $E_{\rm bind,all}$ as a function fo $E_{\rm bind,all}/E_{\rm
bind,b}$ on the right, where $E_{\rm bind,b}$ is an estimate of the
total bulge binding energy assuming a \citet{hernquist1990} profile.
When we use the correlation for major mergers, there is a clear trend
in the residuals -- a $\sim 5\sigma$ significant correlation --
towards an overprediction of the final SMBH mass when the bulge mass is small; $E_{\rm bind,all}$
is not a good predictor of the SMBH for systems with significant disk
contribution to the overall binding energy as compared to $E_{\rm
bind,bulge}$.

The question then arises of why, even in minor merger remnants where it contributes a small fraction of the total mass of the remnant, does the SMBH scale so tightly with the binding energy of the spheroidal component of the potential.  There are two possible scenarios that may explain this correlation: (1) the bulge confines the gas inflow isotropically so that the SMBH can continue to accrete without ejecting gas along potential minima, or (2) the gas content of the final starburst dominates the local potential near the SMBH and therefore it grows until feedback is sufficient to unbind this local gas supply and terminate the inflow.  While we defer a more thorough analysis of these two effects to future work, here we present some suggestive evidence for the latter over the former.

\begin{figure}
\plotone{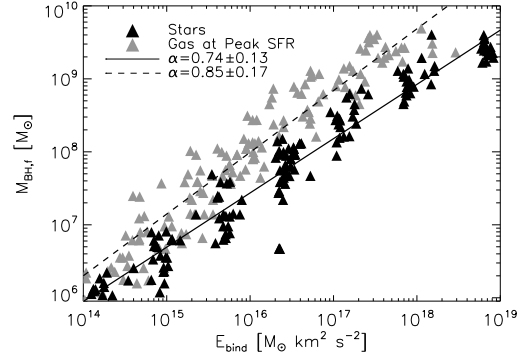}
\caption{From the major mergers series: a comparison of the scaling of the final SMBH mass with the binding energy of the stellar bulge particles after the remnant reaches a state of approximate dynamical equilibrium (black triangles) and gas particles at the peak of the starburst (grey triangles), along with power--law fits to each (solid and dashed lines respectively).  We find that the slope of the gas binding energy is steeper than that of the stellar particles, indicating that the relationship between SMBH mass and bulge binding energy reflects the SMBH self--regulating its growth by terminating the gas inflow, and that gas forming the bulge stars that dominate the potential close to the SMBH.}
\label{fig:bind_compare}
\end{figure}

In Figure~\ref{fig:bind_compare} we present the SMBH scaling with two different binding energy measures: that of the stellar component (the ``stellar binding energy") after the remnant has relaxed, and that of the gaseous component (the ``gas binding energy") at the peak of the final starburst.  We furthermore restrict ourselves to the major merger series, for which it is a fair approximation that the remnant -- including both the stars formed during the simulation and initialized with the progenitor disks -- is entirely bulge dominated and largely gas free.  We find that the relationship between the SMBH mass and the gaseous binding energy is somewhat steeper than that with the stellar binding energy: the logarithmic slopes are $\alpha = 0.85\pm0.17$ and $0.74\pm 0.13$ respectively.  Moreover, this scaling with gas binding energy is statistically consistent with a linear relationship.  This suggests -- albeit tentatively -- that (2) is more likely the dominant process.  However, since the slope is not quite linear, it also argues for some contribution from the confinement of gas by the spheroidal potential.

\section{Conclusions}

We present a series of simulations of the self--regulated growth of
SMBHs in which Eddington--limited accretion continues until a critical
accretion rate, at which point feedback is sufficient to terminate the gas
inflow and cut off further growth.  Simulations of three different
fueling mechanisms -- minor mergers, major mergers, and disk
instabilities -- all follow the same BHFP and binding energy
correlations observed locally \citep{hopkins2007theory,aller2007}.
Increasing the gas content of the initial disks shifts them along
these correlations owing to structural changes in the resulting bulge
arising from the effects of gas dissipation, but does not change their
character.  And, while the major and minor mergers both follow the
same projected correlations -- the $M_{BH}-\sigma$ and
\citet{magorrian1998} relations -- the unstable disks lie along
different correlations, in agreement with observations of pseudobulges
and bulges in barred systems \citep{hu2008,graham2008}, again
reflecting the structural differences between bulges formed via
mergers and secular processes \citep[see also][]{kormendy2004}.  Taken
together, these simulations support the BHFP as the most
``fundamental" scaling relation, in that it reflects the physical
mechanism driving the co-evolution of SMBHs and bulges.

\acknowledgements

Thanks in particular to the anonymous referee for their helpful comments on this manuscript.  We also thank Marijn Franx, Dusan Keres, Stephanie Bush, Gurtina Besla, Chris Hayward, Nick Scoville, Dave Sanders, Lisa Kewley, and Doug Richstone for helpful comments and discussions.  These simulations were performed at the Harvard Institute for Theory and Computation.  This work is supported in part by a grant from the W. M. Keck Foundation.

\bibliographystyle{apj}
\bibliography{../../smbh}

\end{document}